\documentclass[12pt,letterpaper]{article}
\usepackage{amsfonts,amsbsy}

\makeatletter

\renewcommand\section{\@startsection{section}{1}{\z@}%
                                    {-7ex \@plus -1ex \@minus -.2ex}%
                                    {2.5ex \@plus.2ex}%
                                    {\normalfont\large\scshape\centering}}                                   
\renewcommand\subsection{\@startsection{subsection}{2}{\z@}%
                                       {-5ex \@plus -1ex \@minus -.2ex}%
                                       {1.5ex \@plus.2ex}%
                                       {\normalfont\normalsize\scshape}}
\renewcommand\abstract{\section*{\abstractname}}
\newcommand\ack{\section*{\ackname}}       		
\newcommand\ackname{Acknowledgements}
\newcommand\sectionname{}				

\renewcommand\@seccntformat[1]{\ignorespaces\csname #1name\endcsname\space
                               \csname the#1\endcsname.\quad} 	
\renewcommand\appendix{\par
  \setcounter{section}{0}
  \setcounter{subsection}{0}%
  \renewcommand\thesection{\@Alph\c@section}
  \renewcommand\sectionname{\appendixname}}		

\makeatletter
  \divide\@tempdima\baselineskip
  \@tempcnta=\@tempdima
\makeatother

\catcode`\@=11 

\global\newcount\secno \global\secno=0
\global\newcount\meqno \global\meqno=1

\def\newsec#1{\global\advance\secno by1
\global\subsecno=0\eqnres@t
\section{#1}}
\def\eqnres@t{\xdef\secsym{\the\secno.}\global\meqno=1}
\def\sequentialequations{\def\eqnres@t{\bigbreak}}\xdef\secsym{}
\global\newcount\subsecno \global\subsecno=0
\def\subsec#1{\global\advance\subsecno by1
\subsection{#1}}

\def\draftmode{\message{ DRAFTMODE }
\writelabels
 {\count255=\time\divide\count255 by 60 \xdef\hourmin{\number\count255}
  \multiply\count255 by-60\advance\count255 by\time
  \xdef\hourmin{\hourmin:\ifnum\count255<10 0\fi\the\count255}}}
\def\nolabels{\def\wrlabeL##1{}\def\eqlabeL##1{}\def\reflabeL##1{}}
\def\writelabels{\def\wrlabeL##1{\leavevmode\vadjust{\rlap{\smash%
{\line{{\escapechar=` \hfill\rlap{\tt\hskip.03in\string##1}}}}}}}%
\def\eqlabeL##1{{\escapechar-1\rlap{\tt\hskip.05in\string##1}}}%
\def\reflabeL##1{\noexpand\llap{\noexpand\sevenrm\string\string\string##1}}}
\nolabels

\def\eqn#1#2{
\xdef #1{(\secsym\the\meqno)}
\global\advance\meqno by1
$$#2\eqno#1\eqlabeL#1
$$}

\def\eqalign#1{\null\,\vcenter{\openup\jot\m@th
  \ialign{\strut\hfil$\displaystyle{##}$&$\displaystyle{{}##}$\hfil
      \crcr#1\crcr}}\,}

\def\foot#1{\footnote{#1}} 

\global\newcount\refno \global\refno=1
\newwrite\rfile
\def\ref{[\the\refno]\nref}
\def\nref#1{\xdef#1{[\the\refno]}
\ifnum\refno=1\immediate\openout\rfile=refs.tmp\fi
\global\advance\refno by1\chardef\wfile=\rfile\immediate
\write\rfile{\noexpand\bibitem{\string#1}}\findarg}
\def\findarg#1#{\begingroup\obeylines\newlinechar=`\^^M\pass@rg}
{\obeylines\gdef\pass@rg#1{\writ@line\relax #1^^M\hbox{}^^M}%
\gdef\writ@line#1^^M{\expandafter\toks0\expandafter{\striprel@x #1}%
\edef\next{\the\toks0}\ifx\next\em@rk\let\next=\endgroup\else\ifx\next\empty%
\else\immediate\write\wfile{\the\toks0}\fi\let\next=\writ@line\fi\next\relax}}
\def\striprel@x#1{} \def\em@rk{\hbox{}} 
\def\lref{\begingroup\obeylines\lr@f}
\def\lr@f#1#2{\gdef#1{\ref#1{#2}}\endgroup\unskip}
\def\semi{;\hfil\break}
\def\addref#1{\immediate\write\rfile{\noexpand\item{}#1}} 
\def\startrefs#1{\immediate\openout\rfile=refs.tmp\refno=#1}
\def\xref{\expandafter\xr@f}\def\xr@f[#1]{#1}
\def\refs#1{\count255=1[\r@fs #1{\hbox{}}]}
\def\r@fs#1{\ifx\und@fined#1\message{reflabel \string#1 is undefined.}%
\nref#1{need to supply reference \string#1.}\fi%
\vphantom{\hphantom{#1}}\edef\next{#1}\ifx\next\em@rk\def\next{}%
\else\ifx\next#1\ifodd\count255\relax\xref#1\count255=0\fi%
\else#1\count255=1\fi\let\next=\r@fs\fi\next}
\newwrite\lfile
{\escapechar-1\xdef\leftbracket{\string\{}
\xdef\rightbracket{\string\}}}

\def\writestop{\def\writestoppt{\immediate\write\lfile{\string\pageno%
\the\pageno\string\startrefs\leftbracket\the\refno\rightbracket%
\string\def\string\secsym\leftbracket\secsym\rightbracket%
\string\secno\the\secno\string\meqno\the\meqno}\immediate\closeout\lfile}}
\def\writestoppt{}\def\writedef#1{}

\catcode`\@=12 
\def\inv{^{\raise.15ex\hbox{${\scriptscriptstyle -}$}\kern-.05em 1}}

\def\Dsl{\,\raise.15ex\hbox{/}\mkern-13.5mu D} 
\def\dsl{\raise.15ex\hbox{/}\kern-.57em\partial}

\def\tr{{\rm tr}} 

\def\lspace{\ifx\answ\bigans{}\else\qquad\fi}
\def\lbspace{\ifx\answ\bigans{}\else\hskip-.2in\fi} 
\def\boxeqn#1{\vcenter{\vbox{\hrule\hbox{\vrule\kern3pt\vbox{\kern3pt
	\hbox{${\displaystyle #1}$}\kern3pt}\kern3pt\vrule}\hrule}}}
\def\tilde{\widetilde} \def\bar{\overline} \def\hat{\widehat}
\def\e#1{{\rm e}^{^{\textstyle#1}}}

\def\darr#1{\raise1.5ex\hbox{$\leftrightarrow$}\mkern-16.5mu #1}

\def\roughly#1{\raise.3ex\hbox{$#1$\kern-.75em\lower1ex\hbox{$\sim$}}}

\DeclareMathAlphabet{\mathib}{OT1}{cmr}{bx}{it}

\def\a{\alpha}
\def\b{\beta}
\def\c{\chi}
\def\d{\delta}		
\def\e{\varepsilon}	
		
\def\g{\gamma}		
\def\k{\kappa}
\def\l{\lambda}		\def\L{\Lambda} 
\def\m{\mu}

\def\r{\rho}

\def\o{\omega}		\def\O{\Omega}
\def\p{\psi}		
\def\s{\sigma}		\def\S{\Sigma}
\def\th{\theta}

\def\z{\zeta}
\def\CA{{\cal A}}

\def\CC{{\cal C}}

\def\CE{{\cal E}}
\def\CF{{\cal F}}
\def\CG{{\cal G}}

\def\CK{{\cal K}}
\def\CL{{\cal L}}
\def\CM{{\cal M}}
\def\CN{{\cal N}}
\def\CO{{\cal O}}

\def\CV{{\cal V}}
\def\CW{{\cal W}}
\def\CX{{\cal X}}

\def\EJ{\mathfrak{J}}
\def\EM{\mathfrak{M}}
\def\ES{\mathfrak{S}}
\def\BB{\mathib{B}}
\def\BC{\mathib{C}}
\def\BH{\mathib{H}}

\def\C{\mathbb{C}}
\def\rd{\partial}

\def\darr#1{\raise1.5ex\hbox{$\leftrightarrow$}
\mkern-16.5mu #1}

\def\Fr#1#2{{#1\over#2}}

\def\roughly#1{\raise.3ex\hbox{$#1$\kern-.75em
\lower1ex\hbox{$\sim$}}}
\def\ato#1{{\buildrel #1\over\longrightarrow}}

\def\opname#1{\mathop{\kern0pt{\rm #1}}\nolimits}
\def\tr{\opname{Tr}}

\def\End{\opname{End}}
\def\dim{\opname{dim}}
\def\vol{\opname{vol}}
\def\group#1{\opname{#1}}
\def\SU{\group{SU}}
\def\U{\group{U}}

\def\pr{\prime}

\def\bs{\mathib{s}}
\def\bbs{\bar\mathib{s}}
\def\Dp{\rd_{\!A}}
\def\Dpp{\bar\rd_{\!A}}
\def\Da{d_{\!A}}
\def\bari{\bar\imath}
\def\barj{\bar\jmath}
\def\mapr{\!\smash{\mathop{\longrightarrow}\limits^{\bs_+}}\!}
\def\mapl{\!\smash{\mathop{\longleftarrow}\limits^{\bs_-}}\!}

\def\mapd{\Big\downarrow\rlap{$\vcenter{\hbox{$\scriptstyle\bbs_-$}}$}}
\def\mapu{\Big\uparrow\rlap{$\vcenter{\hbox{$\scriptstyle\bbs_+$}}$}}
\def\maprd{\rlap{\lower.3ex\hbox{$\scriptstyle\bs_+$}}\searrow}
\def\mapld{\swarrow\!\!\!\rlap{\lower.3ex\hbox{$\scriptstyle\bs_-$}}}

\def\git{/\kern-.25em/}

\def\cmp#1#2#3{Comm.\ Math.\ Phys.\ {{\bf #1}} {(#2)} {#3}}

\def\np#1#2#3{Nucl.\ Phys.\ {{\bf #1}} {(#2)} {#3}}

\def\top#1#2#3{Topology {{\bf #1}} {(#2)} {#3}}

\def\ivm#1#2#3{Invent.\ Math.\ {{\bf #1}} {(#2)} {#3}}

\makeatother

\lref\SWinv{
E.~Witten,
{\it Monoples and four manifolds},
Math.~Research Lett.~{\bf 1} (1994) 769.
{\tt hep-th/9411102}.
}

\lref\Donaldson{
S.K.~Donaldson,
{\it Polynomial invariants for smooth $4$-manifolds},
\top{29}{1990}{257}.
}

\lref\DK{
S.K.~Donaldson and P.B.~Kronheimer,
{\it The geometry of four-manifolds},
Clarendon Press, Oxford 1990.
}

\lref\SWa{
N.~Seiberg and E.~Witten,
{\it Electric-magenectic duality, monopole condensation, and confinement
in $N=2$ supersymmetric Yang-Mills theory},
\np {B426}{1994}{19},
{\tt hep-th/9411149}.
}

\lref\HPB{
C.~Hofman and J.-S.~Park,
{\it Sigma models for bundles on Calabi-Yau: a proposal for
matrix string compactifications},
Nucl.~Phys.~{\bf B561} (1999) 125-156,
{\tt hep-th/9904150}.
}

\lref\Park{
J.-S.~Park,
{\it N=2 topological Yang-Mills theory
on compact K\"{a}hler surfaces},
\cmp {163}{1994}{113},
{\tt hep-th/9304060}\semi
{\it Holomorphic Yang-Mills theory on compact K\"{a}hler manifolds},
Nucl.~Phys.~ {\bf B423} (1994) 559,
{\tt hep-th/9305095}. 
}

\lref\DT{
S.K.~Donaldson and R.P.~Thomas,
{\it Gauge theory in higher dimensions}, preprint.
}

\lref\ThomasA{
R.P.~Thomas,
{\it A holomorphic Casson invariant for Calabi-Yau $3$-folds, and
bundles on $K3$ fibrations}, 
{\tt math.AG/9806111}.
}

\lref\ThomasB{
R.P.~Thomas,
{\it Gauge theory on Calabi-Yau manifolds},
Ph.D.~thesis, Oxford, 1997.
}

\lref\FukayaC{
K.~Fukaya,
{\it Lagrangian submanifolds and mirror symmetry},
in {Symplectic geometry, mirror symmetry and string theory}, 
1999, KIAS Lecture Notes.
}

\lref\BTE{
M.~Blau and G.~Thompson,
{\it Euclidean SYM theories by time reduction and
special holonomy manifolds},
Phys.~Lett.~{\bf B415} (1997) 242-252,
{\tt hep-th/9706225}.
}

\lref\LLN{
L.~Baulieu, A.~Losev, and N.~Nekrasov
{\it Chern-Simons and twisted supersymmetry in higher
dimensions},
\np {B522}{1998}{82},
{\tt hep-th/9707174}.
}

\lref\BKS{
L.~Baulieu, H.~Kanno, and I.M.~Singer,
{\it Special quantum field theories in eight and other dimensions},
\cmp {194}{1998}{149},
{\tt hep-th/9704167}.
}

\lref\FIM{
J.M.~Figueroa-O'Farrill, A.~Imaanpur, and J.~McCarthy,
{\it Supersymmetry and gauge theory in Calabi-Yau $3$-folds},
Phys.~Lett.~{\bf B419} (1998) 167-174,
{\tt hep-th/9709178}.
}

\lref\ParkB{
J.-S.~Park, 
{\it Cohomological field theories with K\"ahler structure},
{\tt hep-th/9910209}.
}

\lref\Kobayashi{
S.~Kobayashi,
{\it 
Differential geometry of complex vector bundles},
Iwanami Shoten, Publishers and Princeton Univ.~Press,
1987.
}

\lref\HPE{
C.~Hofman and J.-S.~Park,
{\it Holomorphic Casson, Floer and Donaldson-Witten theories},
preprint.
}

\lref\FKO{
K.~Fukaya, M.~Kontsevich, Y.~Oh, H.~Ohta, and K.~Ono,
{\it Anomaly in lagrangian intersection Floer homology},
to appear.
}

\lref\AHS{
M.F.~Atiyah, N.J.~Hitchin and I.M.~Singer,
{\it Self-duality in four dimensional Riemannian
geometry},
Proc.~Roy.~Soc.~London {\bf 362} (1978) {425}.
}

\lref\TFT{
E.~Witten,
{\it Topological quantum field theory},
Comm.~Math.~Phys.~{\bf 117} (1988) 353.
}

\lref\Itoh{
M.~Itoh,
{\it Geometry of Yang-Mills connections over K\"{a}hler
surface},
Proc.~Japan Acad.~{\bf 59} (1983) 431.
}

\lref\Kim{
H.J.~Kim,
{\it Curvatures and holomorphic bundles},
Ph.D.~thesis, Berkely, 1995.
}

\lref\FukayaC{
K.~Fukaya,
{\it Lagrangian submanifolds and mirror symmetry},
in {Symplectic geometry, mirror symmetry and string theory}.
1999.~KIAS Lecture Notes.
}

\lref\tdYM{
E.~Witten,
{\it Two dimensional gauge theories revisited},
J.~Geom.~Phys.~{\bf 9},
 (1992) 303-368,
{\tt hep-th/9204084}.
}

\lref\VW{
C.~Vafa and E.~Witten,
{\it A strong coupling test of S-duality},
Nucl.~Phys.~{\bf B431} (1994) 3-77,
{\tt hep-th/9408074}.
}

\lref\DH{
J.J.~Duistermaat and G.J.~Heckman,
{\it On the variation in the cohomology of the
symplectic form of the reduced phase space},
\ivm {69}{1982}{259}; Addendum, \ivm {72}{1983}{153}.
}

\lref\Wu{
S.~Wu,
{\it An integration formula for the square of momentum
maps of circle actions},
Lett.~Math.~Phys.~{\bf 29} (1993) {311}.
}

\lref\BJSV{
M.~Bershadsky, A.~Johansen, V.~Sadov, and C.~Vafa,
{\it Topological reduction of 4D SYM to 2D $\sigma $--models},
Nucl.~Phys.~{\bf B448} (1995) 166
{\tt hep-th/9501096}.
}

\lref\DPS{
R.~Dijkgraaf, B.J.~Schroers, and J.-S.~Park,
{\it N=4 supersymmetric Yang-Mills theory on a K\"{a}hler surface},
{\tt hep-th/9801066}.
}

\lref\matrixstring{
R.~Dijkgraaf, E.~Verlinde, and H.~Verlinde,
{\it Matrix string theory},
Nucl.~Phys.~{\bf B500} (1997) 43-61,
{\tt hep-th/9703030}.
}

\lref\QMWA{
B.S. Acharya, M.~O'Loughlin, and B. Spence
{\it Higher dimensional analogues of Donaldson-Witten theory}
\np{B503 }{1997}{657},
{\tt hep-th/9705138}.
}

\lref\QMWB{
B.S.~Acharya, J.M.~Figueroa-O'Farrill, M.~O'Loughlin,
and B.~Spence,
{\it Euclidean D-branes and higher-dimensional gauge
theory},
\np{B 514}{1998}{583},
{\tt hep-th/9707118}.
}

\lref\LaL{
J.M.F.~Labastida and C.~Lozano,
{\it 
Mass perturbations in 
twisted N=4 supersymmetric gauge
theories},
Nucl.~Phys.~{\bf B 518} (1998) 37,
{\tt hep-th/9711132}\semi 
{\it 
Duality in twisted N=4 supersymmetric gauge theories in four
dimensions},
Nucl.~Phys.~{\bf B 537} (1999) 203,
{\tt hep-th/9806032}.
}

\lref\BS{
T.~Banks and N.~Seiberg,
{\it Strings from matrices},
Nucl.~Phys.~{\bf B497} (1997) 41-55,
{\tt hep-th/9702187}.
}

\lref\Motl{
L.~Motl,
{\it  Proposals on nonperturbative superstring interactions},
{\tt hep-th/9701025}.
}

\lref\BSV{ 
M.~Bershadsky, V.~Sadov, and C.~Vafa, 
{\it D-branes and topological field theories,} 
{\tt hep-th/9511222}.
}

\lref\TIAN{
G.~Tian, 
{\it Gauge theory and calibrated geometry. I.} 
Ann. of Math. (2) {\bf 151} (2000), no. 1, 193--268.  
}

\begin{document}

\begin{flushright}\scshape
SPIN-1999/26, ITFA-99-30, RUNHETC-2000-37, CU-TP-957\\ 
hep-th/0010103\\
October 2000
\end{flushright}
\vskip1cm

\begin{center}
\renewcommand\thefootnote{\fnsymbol{footnote}}

{\LARGE\scshape Cohomological Yang-Mills Theories\par
on K\"{a}hler 3-Folds \par}
\vskip2cm

{\scshape Christiaan Hofman$^{1,2,\dagger}$ \textnormal{and} 
Jae-Suk Park$^{3,4, \ddagger}$}\\[3mm]

{\itshape 
$^1$New High Energy Theory Center, Rutgers University\\
    126 Frelinghuysen Road, Piscataway NJ 08854, 
    USA\footnote{Present address.}\\[2mm]

$^2$Spinoza Institute \textnormal{and} 
    Institute for Theoretical Physics, University of Utrecht,\\ 
    Leuvenlaan 4, 3508 TD Utrecht, The Netherlands.\\[2mm]

$^3$Department of Physics, Columbia University,\\
    538 West 120th Street, New York, NY 10027,
    USA$^{\thefootnote}$\\[2mm]

$^4$Institute for Theoretical Physics, University of Amsterdam,\\
    Valckenierstraat 65, 1018 XE Amsterdam, The Netherlands.
}\\[2mm]
    
\texttt{$^\dagger$hofman@physics.rutgers.edu, $^\ddagger$jspark@phys.columbia.edu}\\

\end{center}
\setcounter{footnote}{0}

\abstract

We study topological gauge theories with $N_c=(2,0)$ supersymmetry based 
on stable bundles on general K\"ahler 3-folds. In order to have a theory 
that is well defined and well behaved, we consider a model based on an 
extension of the usual holomorphic bundle by including a holomorphic 3-form.  
The correlation functions of the model describe complex 3-dimensional 
generalizations of Donaldson-Witten type invariants. 
We show that the path integral can be written as a sum of contributions 
from stable bundles and a complex 3-dimensional version of  Seiberg-Witten monopoles.
We study certain deformations of the theory, which allow us to consider 
the situation of reducible connections. We shortly discuss situations 
of reduced holonomy. After dimensional reduction to a K\"ahler 2-fold, 
the theory reduces to Vafa-Witten theory. On a Calabi-Yau 
3-fold, the supersymmetry is enhanced to $N_c=(2,2)$.
This model may be used to describe classical limits of certain 
compactifications of (matrix) string theory. 
\newpage
\setcounter{page}{1}


\newsec{Introduction}

The quantum field theoretic approach \TFT\ to Donaldson's 
invariants of $4$-manifolds \Donaldson\DK\
opened up new horizons in mathematics \SWinv\ through the
quantum properties of the underlying physical theory uncovered by
Seiberg and Witten \SWa. The purpose of this paper is
to examine a natural generalization of Donaldson-Witten theory
to a complex K\"{a}hler $3$-fold.

In \Park\ we considered, among others, 
a natural generalization of the Donaldson-Witten theory on 
a complex K\"{a}hler surface to a complex $d>2$ dimensional K\"{a}hler 
manifold $M$. The path integral of the resulting model was localized to the
moduli space of Einstein-Hermitian connections, or equivalently
the moduli space of stable bundles. However this model had a serious
problem due to the uncontrollable abundance of anti-ghosisare only a 
finite dimensional space of anti-ghost zero-modes. 
For higher dimensional K\"ahler manifolds however, we find an 
infinite dimensional solution space. 
Another problem arises due to zero-modes of the ghosts. These are related 
to the appearance of reducible connections (or strictly semi-stable bundles). 
These can also be found for a K\"ahler surface, i.e. $d=2$, but in that case 
one can always get rid of these zero-modes by changing the metric. 
These zero-modes are responsible for the jumps in the observables as a 
function of the metric. In the case of higher dimensional K\"ahler manifold
the appearance of ghost zero-modes is however much more generic and rigid; 
one can in general not get rid of them by a change in the metric. 

In this paper we resolve these problems by starting 
off where we have failed in \Park. A simple observation is that one has to
introduce additional degrees of freedom to control the anti-ghost
zero-modes. This inductive procedure naturally leads us to 
a natural extension of the moduli space of Einstein-Hermitian
connections or, equivalently, stable bundles. This extension is very close to 
the one considered by \FukayaC\ in the context of homological mirror symmetry. 
It turns out that we have a well-defined model only for the $d\leq 3$ case.
By a deformation of the model we are also able to deal with 
reducible connections. In fact, the important ingredient that makes 
this possible is the equivariant treatment, which we adopt from the start. 
We already anticipated this in \Park\HPB. 
Closely related models have been considered in various papers 
\BKS\QMWA\QMWB\BTE\LLN\FIM\TIAN, largely motivated by a program of
Donaldson and Thomas \DT\ThomasA\ThomasB\ as well as certain world-volume
theories of D-branes \BSV.

We will follow the general approach of defining a cohomological 
field theory with a K\"{a}hler structure, as discussed in \ParkB.
We begin by constructing a well-defined $N_c=(2,0)$
model on a K\"{a}hler $3$-fold. This model gives a concrete
formula for Donaldson-Witten type polynomials, which
is valid regardless of what the properties of the extended 
moduli space are. We also argue, using a $S^1$-symmetry
and the DH integration formula, that Donaldson-Witten type 
invariants may be equivalent to Seiberg-Witten type
invariants on K\"{a}hler $3$-folds. 
For manifolds of special holonomy, the model reduces to various known models. 
The dimensional reduction of the model on a 2-torus, gives rise to 
the $N_c=(2,2)$ Vafa-Witten model on a K\"{a}hler $2$-fold.
Finally we briefly specialize to the Calabi-Yau case.
On a Calabi-Yau $3$-fold the $N_c=(2,0)$ supersymmetry is automatically
enhanced to $N_c=(2,2)$ supersymmetry.  
This $N_c=(2,2)$ model can be obtained by dimensional reduction
of the $N_{ws}=(2,2)$ gauged linear sigma model in $(1+1)$ dimensions
introduced in \HPB. The partition function of the theory can be identified 
with the holomorphic Casson invariant defined in \ThomasA\ThomasB.

\newsec{Preliminaries}

In this section we give some preliminary description of 
supersymmetric models and moduli spaces of stable bundles 
on K\"ahler manifolds.

\subsec{General $N_c=(2,0)$ Models}

First we briefly summarize the general structure and some properties 
of cohomological field theories with a K\"{a}hler structure. 
A more detailed discussion can be found in \ParkB.
As discussed in this reference, a cohomological field theory 
on a K\"ahler manifold can always be identified with a $N_c=(2,0)$ 
supersymmetric gauged sigma model in zero dimensions. Such a sigma 
model is classified by data $((\CX,\varpi), \CG, (\CE,h,\ES,\EJ))$, 
where

\begin{itemize}

\item
$\CX$ is a complex K\"{a}hler manifold with K\"{a}hler
form $\varpi$. $\CX$ is the target space of the sigma model.

\item
$\CG$ is a group acting on $\CX$ with isometries. 
This will be the gauge group of the sigma model.

\item
$\CE$ is a $\CG$-equivariant holomorphic Hermitian vector bundle 
over $\CX$ with a Hermitian structure $h$ and
two mutually orthogonal holomorphic sections $\ES$ and
$\EJ$.\foot{More precisely, $\ES$ is a section of the dual holomorphic 
bundle, so that the notion of orthogonality is canonically defined.}

\end{itemize}

The $N_c=(2,0)$ supersymmetry is generated by 
supercharges $\bs_+$ and $\bbs_+$ satisfying
the following anti-commutation relations:
\eqn\alg{
\bs^2=0,\qquad \{\bs,\bbs\}=-i\phi^a_{++}\CL_a,\qquad
\bbs^2=0,
}
where $\phi=\phi^a T^a$ is a $Lie(\CG)$-valued
scalar and $\CL_a$ denotes the Lie derivative
with respect to the vector field $V_a$ on $\CX$
generating the $\CG$-action. The supercharges
$\bs_+$ and $\bbs_+$ can be identified with the 
holomorphic and anti-holomorphic differentials
of the $\CG$-equivariant cohomology of $\CX$ after
parity change. The theory has two additive quantum numbers 
$(p,q)$ called ghost numbers, such that $\bs_+$ has ghost numbers 
$(1,0)$ and $\bbs_+$ has ghost numbers $(0,1)$. 
They define a grading for the fields and observables in the theory. 

Let us now introduce the various ``fields'' of the model. First, we have 
local holomorphic coordinate fields $X^i$on $\CX$, and their complex 
conjugate fields $X^{\bari}$. They are part of holomorphic multiplets 
$(X^i, \p^i_+)$ (i.e.\ $\bbs_+ X^i=0$), and anti-holomorphic multiplets 
$(X^{\bari},\p^{\bari}_+)$ respectively. Their transformation laws are 
\eqn\are{
\eqalign{
\bs_+ X^i&= i\p^i_+ ,\cr
\bbs_+ X^{i}&= 0 ,\cr
\bs_+ X^{\bari} &=0,\cr
\bbs_+ X^{\bari} &=i\p^{\bari}_+,\cr
}\qquad
\eqalign{
\bs_+\p^i_+ &=0  ,\cr
\bbs_+\p^i_+ &= \phi^a_{++}\CL_a X^i  ,\cr
\bs_+ \p^{\bari}_+ &= \phi^a_{++}\CL_a X^{\bari},\cr
\bbs_+\p^{\bari}_+ &=0.
}
}
Associated with the group $\CG$ we have
the $N_c=(2,0)$ gauge multiplet $(\phi_{--},\eta_-,\bar\eta_-,D)$ and 
the invariant field $\phi_{++}$ taking values in $Lie(\CG)$. 
Their transformation laws are
\eqn\ntj{
\eqalign{
\bs_+ \phi_{--} = i\eta_-,\cr
\bbs_+\phi_{--}=i\bar\eta_-,
}
\qquad
\eqalign{
\bs_+ \eta_-&=0,\cr
\bbs_+\eta_- &=+i D + \Fr{1}{2}[\phi_{++},\phi_{--}],\cr
\bs_+\bar\eta_-&=-iD +\Fr{1}{2}[\phi_{++},\phi_{--}],\cr
\bbs_+\bar\eta_-&=0,
}\qquad \eqalign{
\bs_+\phi_{++}=0,\cr
\bbs_+\phi_{++}=0.
}
}
Associated with the holomorphic sections $\ES_\a(X^i)$
and $\EJ^\a(X^i)$, satisfying 
\eqn\holse{
\bbs_+\ES_\a=0,\qquad \bbs_+\EJ^\a=0,
}
we have Fermi multiplets $(\c^\a_-, H^\a)$ and their conjugate
multiplets $(\c^{\bar \a}_-, H^{\bar\a})$, with the
following transformation laws
\eqn\ferm{
\eqalign{
\bs_+\c^\a_- &= - H^\a  ,\cr
\bbs_+\c^\a_- &=  \EJ^\a(X^i),\cr
\bs_+\c^{\bar \a}_- &= \EJ^{\bar  \a}(X^{\bari}) ,\cr
\bbs_+\c^{\bar \a}_- &= - H^{\bar a},
}\qquad
\eqalign{
\bs_+ H^\a &= 0
,\cr
\bbs_+ H^\a &= -i\phi^a_{++}\CL_a \c^\a_- 
+i\p^{i}_+ \rd_i \EJ^\a(X^j)
,\cr
\bs_+ H^{\bar \a} &=
-i\phi^a_{++}\CL_a \c^{\bar \a}_- 
+i\p^{\bari}_+ \rd_{\bari} \EJ^{\bar \a}(X^{\barj})
,\cr
\bbs_+ H^{\bar \a} &=0.
}
}
Note that the section $\EJ^\a(X^i)$ deforms the usual transformation 
$\bbs_+ \c^\a_-= 0$. The holomorphicity of $\EJ^\a(X^i)$ guarantees the 
consistency of the above transformation laws with the commutation 
relations \alg, since $\bbs_+^2\c^\a_- =\bbs_+\EJ^\a(X^i)=0$. 
The Fermi fields $\c^\a_-$ and $\c^{\bar\a}_-$ will be called anti-ghosts 
(they will have negative ghost numbers).

The action functional of the $N_c=(2,0)$ supersymmetric model can be 
given by the following form
\eqn\tzaction{
\eqalign{
S(\zeta)=& -\bs_+\bbs_+\Bigl(\langle\phi_{--}, \m-\zeta\rangle  
-\bigl\langle\eta_-,\bar\eta_-\bigr\rangle
+\Bigl\langle h_{\a\bar\a}(X^i,X^{\bari})\c^\a_-,\c^{\bar \a}_- \Bigr\rangle
\Bigr)
\cr
&
+ i\bs_+\!\bigl\langle\c^\a_-,\ES_\a(X^i)\bigr\rangle 
+i\bbs_+\!\bigl\langle\c^{\bar \a}_-, \ES_{\bar \a}(X^{\bari})\bigr\rangle,
}
}
where $\bigl\langle\cdot,\cdot\bigr\rangle$ denotes a bi-invariant 
inner product on the Lie algebra $Lie(\CG)$, $\m$ denotes the 
$\CG$-equivariant momentum map\foot{The Hamiltonian 
of the $\CG$-action on $\CX$. Note that the K\"ahler manifold 
$\CX$ is automatically symplectic.}  
$\m:X\rightarrow Lie(\CG)^*$, and $\zeta$ is a constant
taking values in the central elements of $Lie(\CG)$.
The condition that the action functional $S(\zeta)$ has
$N_c=(2,0)$ supersymmetry is
\eqn\arh{
\bbs_+\bigl\langle\chi^\a_-, \ES_\a(X^i) \bigr\rangle 
= \bigl\langle\EJ^\a(X^i), \ES_\a(X^i) \bigr\rangle=0,
}
which motivates the orthogonality of the two sections. 

Expanding the action functional $S$ one finds that
the auxiliary fields $D$, $H^\a$ and $H^{\bar \a}$
can be integrated out by setting
\eqn\ari{
D = \Fr{1}{2} (\m-\zeta),\qquad
H^\a= ih^{\a\bar \b}\ES_{\bar\b}.
}
Then the fixed point theorem of Witten implies that
the path integral reduces to an integral
over the space of solutions
of the following equations,
\eqn\arj{
\eqalign{
\EJ^\a(X^i)=0,\cr
\ES_\a(X^i)=0,\cr
\m - \zeta=0,\cr
}
}
and
\eqn\ark{
\phi^a_{++}\CL_a X^i=0,\qquad
[\phi_{++},\phi_{--}]=0,
}
modulo $\CG$-symmetry.
If $\CG$ acts freely on the solution space of \arj, the 
equations \ark\ can only be solved by setting $\phi_{++}=0$, 
and the path integral reduces to an integral  over the symplectic
quotient $\EM_\zeta$ of 
$\left(\ES_{\a}^{-1}(0)\cap \EJ_\a^{-1}(0)\right)\subset \CX$
by $\CG$, 
\eqn\arka{
\EM_{\zeta} = \left(\m\inv(\zeta) \cap \ES_{\a}^{-1}(0)
\cap\EJ_\a^{-1}(0)
\right)/\CG.
}

An observable of the theory $\hat\CO^{r,s}$ is induced by an 
element $\CO^{r,s}$ of $\CG$-equivariant Dolbeault cohomology
of $\CX$ after parity change. The superscript $(r,s)$
denote the ghost numbers and the degrees respectively. 
A correlation function 
$\left\langle\prod_{m=1}^k \widehat \CO^{r_m, s_m}\right\rangle$
can be  non-vanishing only if
\eqn\urkr{
\sum_{m=1}^k (r_m,s_m) = (\triangle,\triangle),
}
where $\triangle$ is the net ghost number
anomaly due to zero-modes of the fermions 
$(\eta_-, \p^i_+,\c^\a_-)$. We call $\triangle$ the formal
complex dimension of $\EM_{\zeta}$. 

If $\CG$ acts freely on 
$(\ES_{\a}^{-1}(0)\cap\EJ_\a^{-1}(0))\subset \C$, we do not have
zero-modes for the ghosts $\eta_-$. If the holomorphic sections 
$\ES$ and $\EJ$ are generic there are no zero-modes of the 
anti-ghosts $\chi^\a_-$. In this situation $\EM_{\zeta}$ is a smooth 
complex $\triangle$-dimensional  non-linear K\"{a}hler manifold.
For non-generic $\ES$ and $\EJ$ the zero-modes of $\chi^\a_-$
span the fibre of a Hermitian holomorphic bundle $\CV\rightarrow
\EM_\zeta$. We call $\CV$ the anti-ghost bundle.
The correlation function 
$\left\langle\prod _{m=1}^k\widehat \CO^{r_m, s_m}\right\rangle$ 
becomes
\eqn\urkt{
\eqalign{
\left\langle\prod _{m=1}^k\widehat \CO^{r_m, s_m}\right\rangle\!
= \int_{\EM_\zeta} \opname{e}(\CV)\wedge
\widetilde \CO^{r_1, s_1}\wedge\ldots\wedge 
\widetilde \CO^{r_k, s_k},
}
}
where $\opname{e}(\CV)$ denotes the Euler class of $\CV$ and
$\tilde\CO^{r_m,s_m}$ denotes a closed differential form
on $\EM_{\zeta}$,
obtained by $\CO^{r_m,s_m}$ after the restriction and
reduction.  It can be non-vanishing if the condition
\urkr\ holds. This ghost number anomaly is reflected geometrically 
by the fact that $\EM_\zeta$ has complex dimension 
$\triangle+\Fr{1}{2}\opname{rank}(\CV)$, while $\opname{e}(\CV)$ 
is a form of degree 
$(\Fr{1}{2}\opname{rank}(\CV),\Fr{1}{2}\opname{rank}(\CV))$. 
So the integrand of 
the RHS of \urkt\ is a top form exactly if \urkr\ holds.

\subsec{A Target Space from Bundles On K\"ahler Manifolds}

We now describe a $N_c=(2,0)$ model related to stable bundles on a 
K\"ahler manifold. For general references on these structures 
see \DK\Kobayashi.
We consider a compact complex K\"{a}hler $d$-fold $M$ 
with K\"{a}hler form $\o$.  
The complex structure on $M$ determines a decomposition of 
the space $\O^r(M)$ of $r$-form on $M$ as $\O^r(M) = \oplus_{p+q=r}\O^{p,q}(M)$.
On  $M$ any two-form $\a\in \O^2(M)$ 
can be decomposed into 
$\a =\a^+ +\a^-$ such that 
\eqn\bcg{
\eqalign{
\a^+ &= \a^{2,0} + \a_0 \o + \a^{0,2},\cr
\a^-  & = \a^{1,1}_{\perp},
}
}
where $\a_0\in\O^0(M)$ is a scalar function 
and $\a^{1,1}_{\perp}$ is a $(1,1)$-form orthogonal to $\o$.
Corresponding to this decomposition we define the following projections
\eqn\kab{
P^\pm:\O^2(M)\rightarrow \O^{2\pm}(M),\qquad 
P^{0,2}: \O^2(M)\rightarrow \O^{0,2}(M).
}
For a complex K\"{a}hler $2$-fold the above decomposition coincides
with the decomposition in self-dual and anti-self dual two-forms. 
We denote by $\O^p(M, E)$ the space of real $p$-forms on $M$
taking values in $E$.
Let $E$ be a rank $r$ vector bundle over $M$
endowed with a Hermitian metric. The choice of $E$ fixes the topological
type for the connections on $E$.  We denote by $\CA$ the
space of all connections and by $\CG$ the group of all gauge
transformations.
The gauge group $\CG$ is equivalent to the group of all unitary
automorphisms of $E$ (and it has structure group $\U(r)$). 
The Lie algebra $Lie(\CG)$ of $\CG$ can be identified with 
$\O^0(M, \End(E))$ and we use integration over
$M$ to identify $Lie(\CG)^*$ with $\O^{2d}(M, \End(E))$.
Thus the bi-invariant inner product on $Lie(\CG)$ is
the integral over $M$ combined with the trace of $\U(r)$ 
\eqn\bca{
\langle a,a \rangle = -\int_M \tr (a\wedge * a).
}
We take the infinite dimensional space $\CA$ as our initial 
target space $\CX$. (Later in this paper we shall extend this space). 

To define an equivariant $N_c=(2,0)$ model we
need to introduce complex and K\"{a}hler structures
on our target space $\CA$.
Let $A$ denote a connection one-form, which is decomposed into
$A=A^{1,0} + A^{0,1}$.  We denote by $\Da =\Dp +\Dpp$
the corresponding covariant derivative,
\eqn\vaba{
\Da =\Dp +\Dpp: \O^0(M,E) \longrightarrow \O^{1,0}(M,E)
\oplus \O^{0,1}(M, E).
}
The space $\CA$ is an infinite dimensional affine space. 
A tangent vector is represented by $\d A\in\O^1(M, \End(E))$.
Note that there is no natural complex structure on $\CA$. Any complex
structure should be induced from the complex structure on $M$.
One introduces a complex structure on $\CA$ by declaring 
$\d A^{0,1}\in\O^{0,1}(M, \End(E))$ to be the holomorphic
tangent vectors. Then $\CA$ becomes an infinite dimensional
flat K\"{a}hler manifold with K\"{a}hler form $\varpi$ given by 
\eqn\vac{
\varpi(\d A,\d A') =  \Fr{1}{4~d!\pi^2}\int_M \tr (\d A\wedge\d A')\wedge \o^{d-1},
}
on which $\CG$ acts with isometries preserving the K\"ahler structure.
The K\"{a}hler potential for the K\"{a}hler form \vac\ of  $\CA$ is given by
\eqn\bab{
\CK(A^{1,0},A^{0,1}) = 
\Fr{1}{4~d!\pi^2}\int_M \k \tr (F\wedge F)\wedge\o^{d-2},
} 
where $\k$ is a K\"{a}hler potential for $\o$, i.e., $\o= i\rd\bar\rd \k$.

Now we introduce our $N_c=(2,0)$ supercharges $\bs_+$ and
$\bbs_-$ with the familiar commutation relations
\eqn\bcc{
\bs_+^2=0,\qquad \{\bs_+,\bbs_+\} = -i \phi^a_{++}\CL_a,
\qquad \bbs_+^2=0.
}
The supercharges are identified with the differentials of
$\CG$-equivariant cohomology of our target space $\CA$.
Thus $\phi^a_{++}\CL_a$ is the infinitesimal gauge transformation
generated by the adjoint scalar $\phi_{++} \in Lie(\CG) = \O^0(M,\End(E))$.
The $N_c=(2,0)$ gauge multiplet $(\phi_{--},\eta_-,\bar\eta_-,D)$
takes values in $\O^0(M,\End(E))$. Their
transformation laws for are given by the general formula \ntj. 

With the complex structure on $\CA$ introduced above
we have holomorphic multiplets $(A^{0,1}, \p^{0,1}_+)$ and 
conjugate anti-holomorphic multiplets $(A^{1,0},\bar\p^{1,0}_+)$, respectively,
where $\p^{0,1}_+\in \Pi\O^{0,1}(M, \End(E))$ represents a holomorphic
cotangent vector in $\CA$. These are the multiplets associated to the 
coordinates $X^i$ and $X^{\bari}$. The transformation laws are as given 
as in \are\ (or in more details in Appendix A). Note that
\eqn\bcf{
\{\bs_+,\bbs_+\} A = -i d_{\! A}\phi_{++},\qquad
\{\bs_+,\bbs_+\} \p^{0,1}_+ = i [\phi_{++},\p^{0,1}_+],
}
which are the infinitesimal gauge transformations generated
by $\phi_{++}$, in accordance with \bcc. 

From the transformation laws and the K\"ahler form \bab\ we 
obtain the following equivariant K\"{a}hler form
\eqn\bae{
\eqalign{
\widehat\varpi^\CG&=
i\bs_+\bbs_+ \CK\cr
&=\Fr{i}{2~d!\pi^2}\int_M \tr (\phi_{++} F)\wedge \o^{d-1} 
+\Fr{1}{2~ d!\pi^2}
\int_M \tr(\p^{0,1}_+\wedge\bar\p^{1,0}_+)\wedge \o^{d-1},
}
}
where we used the Bianchi identity $\Da F=0$, which implies
$\Dpp F^{0,2}=\Dp F^{0,2}+ \Dpp F^{1,1}=0$, and integration
by parts.  
The second term of the equivariant K\"ahler form can be identified 
with the K\"{a}hler form $\varpi$ (after parity change) and the 
first term is the $\CG$-momentum map $\phi^a_{++}\m_a$,
$\m:\CA\rightarrow Lie(\CG)^*=\O^{2n}(M,\End(E))$,
\eqn\baf{
\m(A) =  \Fr{1}{2~d!\pi^2} F\wedge \o^{d-1}
= \Fr{1}{2d~d!\pi^2}(\L F) \o^d,
}
where $\L$ denotes the adjoint of wedge multiplication with $\o$.

With the construction described until now, we have a $N_c=(2,0)$ model 
based on our infinite dimensional target space $\CA$. 
The path integral of the resulting model will localize 
to the symplectic quotient $\m\inv(\z)/\CG$. 
For $d \geq 2$ the quotient
space is still infinite dimensional. Thus we should supply some
additional localization. According to our
general discussion in the last section we may still  
consider a certain infinite dimension Hermitian holomorphic
vector bundle $\CE\rightarrow \CA$ over $\CA$
with a certain holomorphic section $\ES$ 
(we will put $\EJ=0$ for the moment), 
which determines anti-ghost multiplets accordingly. 
Then the path integral will be further localized to 
$(\ES\inv(0)\cap\m\inv(\zeta))/\CG$,
which might be a finite dimensional K\"{a}hler manifold.
We will now consider such an extension of the model.

\subsec{The Holomorphic Section}

The remaining task is to determine an infinite dimensional 
vector bundle over our target space $\CA$ with an appropriate
$\CG$-equivariant holomorphic section $\ES(A^{0,1})$, 
i.e.~$\bbs_+\ES=0$. From our general discussion we can see that 
a choice of section $\ES$ should be compatible
with the K\"{a}hler quotient such that the effective target space
$\CM= (\ES\inv(0)\cap \m\inv(\zeta))/\CG$ inherits a K\"{a}hler
structure when $\CG$ acts freely.
We introduce a bundle $\CE$ over our target space $\CA$ for which a
holomorphic section $\ES(A^{0,1})$ is given by 
\eqn\bda{
\ES: A^{0,1}\rightarrow F^{0,2}\in \O^{0,2}(M, \End(E)).
} 
We note that the above is the most natural choice on {\it generic}
K\"{a}hler manifolds, since any holomorphic function  
of $A^{0,1}$ which is gauge covariant must be a function of $F^{0,2}$.
A further obvious requirement is that the resulting action functional
should be invariant under the Lorentz symmetry -- more precisely 
the holonomy of a K\"{a}hler manifold $M$.\foot{There are two special cases.
On  a Calabi-Yau $4$-fold or an arbitrary hyper-K\"{a}hler
manifold one can take a certain projection of $F^{0,2}$ for the holomorphic
section of $\CE\rightarrow \CA$. We will return to this in another paper\HPE.}
Then our effective target space will be the  moduli space
$\CM_{EH}$ of Einstein-Hermitian bundles defined by
\eqn\bde{
\CM_{EH} = (\ES\inv(0)\cap \m\inv(\zeta))/\CG.
}
Since our section takes values in $\O^{0,2}(M, \End(E))$
we have corresponding Fermi multiplets $(\chi^{2,0}_-, H^{2,0})$,
taking values in $\O^{2,0}(M, \End(E))$. They transform according 
to the general transformation laws \ferm, with $\EJ=0$. 

Now we have all the ingredientsnecessary to define a $N_c=(2,0)$ model. 
For example, the action can be found from the general form \tzaction.

\newsec{Motivating The Extended Moduli Space Of Stable Bundles}

In this section we motivate the notion of extended moduli space
of stable bundles \FKO, in the context of resolving the problems of 
anti-ghost zero-modes.

First we set up our notation.
Consider a $d$ complex dimensional compact
K\"{a}hler manifold $(M,\o)$ with K\"{a}hler form
$\o$, and a rank $r$ Hermitian vector bundle $E\rightarrow M$.
The curvature two-form decomposes as $F = F^+ + F^-$ according to \bcg. 
A connection on $E$ is called Einstein-Hermitian (EH) 
with factor $\zeta$ if
\eqn\kac{
\eqalign{
F^{0,2} &= 0,\cr
i\L F &= \zeta I_{E}.
}
}

The model as it now stands has a problem with the anti-ghost zero-modes.
Let $A$ be an EH connection. We consider a nearby connection
$A +\d A$, $\d A\in \O^1(M,\End(E))$, which also is EH. After linearization
we have $d_{\! A}^+\d := P^+d_{\! A}\d A = 0$, with $P^+$ the projection 
operator defined in \kab. There is still a gauge freedom $d_A\lambda$.
Supplying the Coulomb gauge condition 
$d_{\!A}^* \d A=0$, local deformations $\d A$ around a point $A$ 
in $\CM_{EH}$ are represented by the kernel of the operator 
$d^+_{\!A}\oplus d^*_{\!A}$ acting on $\O^1(M,\End(E))$.
This structure can be summarized by the associated elliptic 
complex of Atiyah-Hitchin-Singer \AHS;
\eqn\kad{
0\longrightarrow
\O^0(M,\End(E)) \,\ato{d_{\!A}}\,
\O^{1}(M,\End(E))\, \ato{d_{\!A}^+}\,
\O^{2+}(M, \End(E)).
}
We compare this complex with the fermionic zero-modes
of the fermions $(\bar\eta_-, \p^{0,1}_+,\bar\chi^{0,2}_-)$ 
in the model introduced in Sect.~2.2, which are governed by the equations
\eqn\bdg{
\Dpp \bar\eta_- =0,\qquad
\eqalign{
\Dpp^* \p^{0,1}_+ =0,\cr
\Dpp \p^{0,1}_+=0,\cr
}\qquad
\Dpp^*\bar\chi^{0,2}_-=0.
}
After decomposing $\bar\eta_- = \boldsymbol{\eta}_- + i\c^0_-$
into its real and imaginary part, we can form
real fermions $(\boldsymbol{\eta}_-, 
\boldsymbol{\psi}_+,\boldsymbol{\chi}_-)$
which we define as 
\eqn\kae{
\boldsymbol{\psi}_+ =\bar\p^{1,0}_+ + \p^{0,1}_+,\qquad
\boldsymbol{\chi}_- =\bar\c^{2,0}_- +   \chi^0_-\o + \c^{0,2}_-,
}
so that $\boldsymbol{\eta}_-\in \O^0(M,\End(E))$, 
$\boldsymbol{\psi}_+\in\O^1(M,\End(E))$ and
$\boldsymbol{\chi}_- \in \O^{2+}(M,\End(E))$.
The zero-mode equations \bdg\ are then translated into 
\eqn\kaf{
\Da \boldsymbol{\eta}_-=0,\qquad
\eqalign{
\Da^*\boldsymbol{\psi}_+=0,\cr
\Da^+\boldsymbol{\psi}_+=0,\cr
}\qquad
\Da^{+*}\boldsymbol{\chi}_-=0.
}
Thus the zero-modes of the fermions $(\boldsymbol{\eta}_-, 
\boldsymbol{\psi}_+,\boldsymbol{\chi}_-)$
are elements of the AHS complex \kad.
The above correspondence is one of the crucial
ingredients of Witten's approach to 
Donaldson theory in four real dimensions \TFT.
The path integral measure contains such fermionic
zero-modes and the net ghost number anomaly is
precisely the index of the above complex, which is
the formal dimension of the moduli space of instantons
on a four manifold. 

Let us undo the combination \kae, and return to
the initial equations \bdg\ for the complex fermions
$(\bar\eta_-, \p^{0,1}_+, \bar\chi^{0,2}_-)$.
The equations \bdg\ imply that the fermionic
zero-modes are in one to one correspondence
with the following Dolbeault complex \Itoh
\eqn\kag{
0\longrightarrow
\O^{0,0}(M, \End(E))\, \ato{\Dpp}\,
\O^{0,1}(M, \End(E))\, \ato{\Dpp}\,
\O^{0,2}(M, \End(E)).
}
Note that $\Dpp^2=0$ at the fixed point locus.
Our problem for $d\geq 3$ is that a fermionic
zero-mode of $\bar\chi^{0,2}_-$  only needs to satisfy
the condition $\Dpp^*\bar\chi^{0,2}_-=0$. As a result
we always have an infinite dimensional anti-ghost bundle.
Therefore the path integral would hardly make any sense.
But this is exactly what the EH condition gives us
via local deformations.
For $d=2$ the desired condition $\Dpp\bar\chi^{0,2}_-=0$
is void due to the dimensional reason.
For $d\geq 3$ the only way of imposing the desired
condition $\Dpp\bar\chi^{0,2}_-=0$ is to introduce
another fermionic field $\l_+^{3,0}$ with ghost numbers
$(1,0)$ such that the action functional contains
the following term
\eqn\kga{
S \sim \int_M\tr(\l_+^{3,0}\wedge * \Dpp\bar\chi^{0,2}_-) +\cdots.
}
Then we obtain in addition to \bdg\ the two equations 
\eqn\kgb{
\Dpp\bar\chi^{0,2}_- =0,\qquad
\Dp^*\l_+^{3,0} =0.
}
Thus we have to generalize the $N_c=(2,0)$ model
by introducing a new holomorphic multiplet
$(C^{3,0}, \l_+^{3,0})\in \O^{3,0}(M, \End(E))$.\foot{
When $\l^{3,0}_+$ is in a Fermi multiplet it is impossible to 
get the term \kga\ without breaking the $N_c=(2,0)$ supersymmetry.}
For $d=3$ the above additional conditions are sufficient.
For $d\geq4$ we should supply yet another additional
condition $\Dp\l^{3,0}_+=0$, otherwise we have
too many zero-modes for $\l^{3,0}_+$. Thus we should
introduce another fermionic field $\bar\xi^{0,4}_-$
with ghost numbers $(-1,0)$ such that now the action contains 
\eqn\kgga{
S \sim \int_M\tr(\l_+^{3,0}\wedge * \Dpp\bar\chi^{0,2}_-
+ \Dp\l_+^{3,0}\wedge * \bar\xi^{0,4}_-) +\cdots,
}
and so on.

Thus a natural resolution of our problem is to extend
the complex \kag\ all the way to the end
\eqn\kah{
0\longrightarrow
\BC^{0,0}\, \ato{\Dpp}\,
\BC^{0,1}\, \ato{\Dpp}\,
\BC^{0,2}\, \ato{\Dpp}\,
\BC^{0,3}\,\ato{\Dpp}\,
\ldots\,\ato{\Dpp}\,
\BC^{0,d}\longrightarrow 0,
}
where $\BC^{0,\ell}:=\O^{0,\ell}(M,\End(E))$.
To give any meaning to the above Dolbeault complex,
we have to introduce the following
set of fermionic fields
\eqn\kai{
\bar\eta^{0,0}_-, \psi^{0,1}_+, \bar\chi^{0,2}_-,\bar\l^{0, odd}_+,
\bar\xi^{0,even}_-, 
}
where $2 < odd, even \leq d$. 
It can be seen, from the basic structure of our $N_c=(2,0)$ model,
that $\bar\l^{0,odd}_+$ are superpartners of anti-holomorphic 
bosonic fields  $C^{0,odd}$, forming anti-holomorphic multiplets;
\eqn\kba{
C^{0,odd}\, \ato{\bbs_+}\, \bar\l^{0,odd}_+.
}
Furthermore, the fields 
$\bar\xi^{0,even}_-$ should be in Fermi multiplets
\eqn\kbb{
\bar\xi^{0,even}_-\,\ato{\bs_+}\, H^{0,even},
} 
where $H^{0,even}$ are auxiliary fields. 
Then we may try to design an action functional
which gives the following equations, in addition 
to \bdg, for fermionic zero-modes
\eqn\kaj{
\eqalign{
\Dpp \bar\l^{0,odd}_+ =0,\cr
\Dpp^*\bar \l^{0,odd}_+=0,\cr
}
\qquad
\eqalign{
\Dpp \bar\xi^{0,even}_- =0,\cr
\Dpp^*\bar\xi^{0,even}_-=0.\cr
}
}
Thus the $(0,q)$-form fermionic zero-modes become the elements
of the $q$-th cohomology group 
$\BH^{0,q}:=H^{0,q}_{\Dpp}(M, \End(E))$
of the complex \kah. 
Then the net ghost number violation due to the fermionic
zero-modes is precisely the index
$\sum_{q=0}^d (-1)^{q+1} \opname{\dim_\mathbb{C}} \BH^{0,q}$
of the complex \kah. Now we are in the same situation
as the Donaldson-Witten theory in the $d=2$ case.

Finally let's consider how the above extension fits into
the framework of EH connections. Kim \Kim\ introduced the 
followingcomplex (see also \Kobayashi), generalizing the 
complex given in \kad,
\eqn\kal{
0\longrightarrow
\BB^0 \,\ato{d_{\!A}}\,
\BB^{1}\, \ato{d_{\!A}^+}\,
\BB^{2+}\, \ato{d_{\!A}^{0,2}}\,
\BB^{0,3}\,
 \ato{\Dpp}\,
\ldots  \ato{\Dpp}\,
\BB^{0,d}
\longrightarrow 0,
}
where $d^{0,2}_{\!} = \Dpp\circ P^{0,2}$,
$\BB^p = \O^p(M, \End(E))$ and
$\BB^{p,q}= \O^{p,q}(M,\End(E))$.
It is shown that the above is a complex if the connection $A$
is EH and elliptic. 
We denote the associated $q$-th cohomology group by
$\BH^q$. It is not difficult to show
that
\eqn\kam{
\sum_{q=0}^d (-1)^{q+1} \opname{\dim_\mathbb{R}}\BH^q 
= 2\sum_{q=0}^d (-1)^{q+1} \opname{\dim_\mathbb{C}}\BH^{0,q}.
}
It should also be obvious that the two extended complexes \kal\
and \kah\ are related in the same way as the unextended 
complexes \kad\ and \kag. 

We remark that Kim's complex is not the genuine deformation
complex of EH connections, but rather a natural extension
of it. 
As in the $d=2$ case we require that the index is 
the formal complex dimension of a certain extended 
moduli space of stable bundles.
We define the extended  moduli space $\EM$
of EH connections or stable bundles 
by extending the EH condition as the space of solutions of the 
following equations 
\eqn\kima{
\eqalign{
\bar\mathfrak{D}\circ\bar\mathfrak{D} =0,\cr
\exp(\o)\wedge\left(\mathfrak{D}\circ\bar\mathfrak{D} 
+ \bar\mathfrak{D}\circ 
\mathfrak{D}\right)|_{\hbox{top form}} + i d\zeta \o^d I_E=0,\cr
}
}
where 
$\bar\mathfrak{D}$ is the extended holomorphic connection
\eqn\kimb{
\bar\mathfrak{D} = \Dpp + \sum_{k\geq 1} C^{0,2k+1}.
}
The versal deformation complex of the above equations 
is then precisely equivalent to Kim's complex \kal. 
This can be checked using the K\"{a}hler identities 
\eqn\khiden{
\Dpp^*=-i[\L,\Dp],\qquad
\Dp^*=i[\L,\Dpp].
}
In the above scheme the infinitesimal deformations
of the extended moduli space always lie in $\BH^{0,odd}$,
while the obstructions, by Kuranishi's method, lie in 
$\BH^{0, even}$. Thus the local model of the extended
moduli space is $f^{-1}(0)$ \FukayaC, where
\eqn\kurani{
f: \BH^{0,odd}\rightarrow\BH^{0,even},\quad
f(A,C)=\bar\mathfrak{D}\circ\bar\mathfrak{D}.
}
The formal complex dimension of the extended moduli space 
$\EM$ can be computed using the Riemann-Roch formula
\eqn\kimc{
\sum_{q=0}^d (-1)^{q+1} \opname{\dim_\mathbb{C}}\BH^{0,q}
=-\int_M \opname{Td}(M)\wedge \opname{ch}(E)\wedge \opname{ch}(E^*),
}
where $\opname{Td}(M)$ denotes the Todd class of $M$ and $\opname{ch}(E)$
denotes the Chern character of $E$.

It seems that we have all the ingredients to construct a well-defined
$N_c=(2,0)$ model. Unfortunately it turns out to be
impossible to implant the above ideas, except for
the case of at most three complex dimensions. 
It is not possible to maintain $N_c=(2,0)$ supersymmetry and impose
the desired equations \kaj\ for all fermions unless $d\leq 3$.
This follows from the fact that the zero-mode equations for 
the fermions in the holomorphic multiplet should be completely 
holomorphic equations (they arise from the supersymmetry 
transformation of the first two equations in \arj). 
This is inconsistent with the two equations for $\l_+^{odd,0}$ 
in \kaj, therefore we can impose at most one of them. This is 
sufficient only for $d\leq 3$. The reason why we did not have 
this problem in lower dimensions was that for $\psi^{1,0}_+$ 
we also had the non-holomorphic supersymmetric partner of 
the D-term equation at our disposal. This equation is 
related to the gauge symmetry. So in order to extend the 
above ideas to higher dimensions, we are led to associate 
the even degree terms in the complex \kal\ with new gauge 
symmetries rather than obstructions. We do however not see 
how these can be related to gauge symmetries, except for 
$\BB^0$. Therefore in the rest of this paper we will only 
consider $d\leq3$.

\newsec{$N_c=(2,0)$ Model On K\"{a}hler 3-Folds}

We consider  the $N_c=(2,0)$ model studied in Sect.~$2$
specializing to the case that $M$ is a  K\"{a}hler $3$-fold. 
According to the discussion in the previous section
we introduce one more bosonic field $C^{0,3}\in \O^{0,3}(M,\End(E))$
and its Hermitian conjugate $C^{3,0}$.
Our goal
is to construct a $\CG$-equivariant $N_c=(2,0)$ model
whose target space is the space $\CA$ of all connections
together with the space of all $C^{0,3}$ fields.
Furthermore the fermionic zero-modes should be 
elements of the Dolbeault cohomology of the complex \kah.
It turns out there is only one way of achieving this goal.

\subsec{Basic Properties Of The Model}

The $N_c=(2,0)$ model here will be an example
of the construction in Sect.~$2.1$ with $\EJ\neq 0$ in \ferm.
We first recall that the path integral of a general $N_c=(2,0)$ model
is localized to the solution space of \arj, modulo $\CG$ symmetry.
The momentum map $\m$ is determined from the K\"{a}hler potential
on the space of all $X^i$ and from the action of $\CG$ on it.
The sections $\EJ^\a$ and $\ES_\a$ should satisfy the following
equations to have $N_c=(2,0)$ supersymmetry,
\eqn\pkb{
\eqalign{
\bbs_+ \EJ^\a =0,\cr
\bbs_+ \ES_\a =0,\cr
\bbs_+ \ES_\a=0,\cr
\left\langle\EJ^\a, \ES_\a\right\rangle=0.\cr
}
}

In the present case our infinite dimensional target space is 
\eqn\pkc{
\CX = \CA \oplus \O^{3,0}(M, \End(E))\oplus \O^{0,3}(M,\End(E)),
}
and the infinite dimensional group $\CG$ acts on the above space
as  the group of all local gauge transformation on $M$. 
The Lie algebra $Lie(\CG)$ of $\CG$
is $\O^0(M, \End(E))$ and the bi-invariant inner product on $Lie(\CG)$
is \bca.
We already gave a complex structure on $\CA$ in Sect.~$2.2$
by demanding that $A^{0,1}$ is a holomorphic field, i.e., 
$\bbs_+ A^{0,1}=0$. We have a unique holomorphic section $F^{0,2}$ 
from the subspace $\CA$ and the corresponding Fermi multiplet 
$(\bar\chi^{0,2}_-, H^{0,2})\in \O^{0,2}(M, \End(E))$ with the 
transformation laws \ferm. Let us see what the complex structure on 
the additional field space should be. We need to put a constraint 
on the additional fields $C^{3,0}$. From the discussion in 
the last section this condition is $\Dp^* C^{3,0}=0$. 
This constraint has to come from either one of the first two 
equations in \arj\ (or their conjugates), and therefore must be 
(anti-)holomorphic. Note that $\Dp^*=-*\Dpp *$, which is holomorphic, 
since $\bbs_+ A^{0,1}=0$. Therefore, for the equation to be holomorphic, 
we need also $\bbs_+ C^{3,0}=0$. Thus the additional holomorphic multiplet is 
$(C^{3,0}, \l^{3,0}_+)$. The additional equation could be added to $\ES$, 
as it has the same form-degree $(0,2)$ (after conjugation), 
so that we get that the combination $F^{0,2}-\Dpp^* C^{0,3}$ has to 
vanish.\foot{This implies the equation
$\Dpp^* C^{0,3}=0$ due to the Bianchi identity. This in fact is the 
combination that is often used in the literature \FukayaC\LLN.}
This is however not possible in our setting, because $F^{0,2}$ is 
holomorphic, while the second part is 
anti-holomorphic, as we just argued. Therefore, the total combination is 
neither holomorphic nor anti-holomorphic, as is required for $\ES$ 
(respectively $\bar \ES$). Therefore, our only choice is to use the 
first equation in \arj, that is we should set $\EJ=\Dp^* C^{3,0}$. 
We see that $\bbs_+\EJ =0$. We conclude
\eqn\pke{
\eqalign{
\EJ &= \Dp^* C^{3,0},\cr
\ES &= F^{0,2}.
}
}
With this choice also the last condition in \pkb\ is satisfied, as 
\eqn\pke{
\langle\EJ^\a,\ES_\a\rangle = 
\int_M\tr(\Dp^* C^{3,0}\wedge * F^{0,2}) =
\int_M\tr(C^{3,0}\wedge *\Dpp F^{0,2}) = 0,
}
where we used the Bianchi identity $\Da F = 0$, which implies $\Dpp F^{0,2}=0$.

The above considerations determine an equivariant $N_c=(2,0)$ model, 
following the description in Sect.~$2.1$.

\subsec{Fields And Action Functional}

Here we recall again the fields and 
their supersymmetry transformation laws, 
to summarize what we have learned.
Associated with the $\CG$ symmetry we have
the $N_c=(2,0)$ gauge multiplet
$(\phi_{--}, \eta_-,\bar\eta_-,D)$, all transforming 
as adjoint valued scalars on $M$.
The transformation
laws are given by \ntj.
We have two sets of holomorphic multiplets and their 
anti-holomorphic partners.
One set of holomorphic multiplets is $(A^{0,1},\p^{0,1}_+)$  
with anti-holomorphic partners $(A^{1,0},\bar\p^{1,0}_+)$.
The other  holomorphic multiplet is $(C^{3,0},\l^{3,0}_+)$ 
with anti-holomorphic partner $(C^{0,3},\bar\l^{0,3}_+)$.
Finally we have Fermi multiplets $(\chi^{2,0}_-, H^{2,0})$
and anti-Fermi multiplets  $(\bar\chi^{0,2}_-, H^{0,2})$.
The explicit transformation rules are written down in Appendix A. 
The fields and their transfoormation rules can be summarized by 
the following diagrams, 
\eqn\pxd{
\def\normalbaselines{\baselineskip20pt
\lineskip3pt \lineskiplimit3pt}
\matrix{
C^{3,0}    & \mapr & \l^{3,0}_+ &  & \phi_{++}      \cr
           &       &            &  &                \cr
\bar\eta_- & \mapr & D          &  & \bar\l^{0,3}_+ \cr
\mapu      &       & \mapu      &  & \mapu          \cr
\phi_{--}  & \mapr & \eta_-     &  & C^{0,3}        \cr
}\quad,\qquad
\matrix{
                 &       & A^{0,1} & \mapr & \p^{0,1}_+ \cr
\cr
\bar\chi^{0,2}_- & \mapr & H^{0,1}
}.
}

The resulting $N_c=(2,0)$ model in general can not
be embedded into a $N_c=(2,2)$ theory since $\bs_+\bar\chi^{0,2}_-\neq 0$.
Such an embedding is only possible if $M$ is a Calabi-Yau $3$-fold,
where our $N_c=(2,0)$ supersymmetry will automatically 
enhance to $N_c=(2,2)$ even without adding additional fields.

The final ingredient for the action functional is the $\CG$-momentum map
on the  total space \pkc.
The total space has a natural $\CG$-invariant K\"{a}hler potential
\eqn\pla{
\CK_T = \Fr{1}{24\pi^2}\int_M \biggl(\k \tr(F\wedge F)\wedge \o^2
-i \tr \left(C^{3,0}\wedge  C^{0,3}\right)\biggr).
}
Using the transformation laws in Appendix A we obtain from this 
the following equivariant K\"{a}hler form,
\eqn\plb{
\eqalign{
\widehat\varpi^\CG_T:=&
i\bs_+\bbs_+ \CK_T \cr
=&\Fr{1}{12\pi^2}\int_M \tr \Bigl(i\phi_{++} 
\Bigl(F\wedge\o^2 +\Fr{1}{2}[C^{3,0},C^{0,3}]\Bigr)\Bigr)
\cr
&
+\Fr{1}{12\pi^2}
\int_M \tr\Bigl(\p^{0,1}_+\wedge\bar\p^{1,0}_+\wedge\o^2
-\Fr{i}{2}\l^{3,0}_+\wedge\bar\l^{0,3}_+
\Bigr).
}
}
The last line is the K\"{a}hler form $\widehat\varpi_T$,
after parity change,
and the term in the second line is proportional to 
the $\CG$-momentum map $\m_{T}$ on the total
space \pkc,
\eqn\plc{
\m_T =  \Fr{1}{12\pi^2} \Bigl(F\wedge\o^2 +\Fr{1}{2}[C^{3,0},C^{0,3}]\Bigr).
}
Thus the $N_c=(2,0)$ action functional
is given by, following \tzaction,
\eqn\tfaxk{
\eqalign{
S = &\Fr{\bs_+\bbs_+}{12\pi^2}\int_{M}\!\tr\biggl(
\phi_{--} \!\biggl(F\wedge\o^2 +\Fr{1}{2}[C^{3,0},C^{0,3}]
+\Fr{i}{3}\zeta \o^3 I_E\biggr)\biggr)
\cr
&
+\Fr{\bs_+\bbs_+}{4\pi^2}\int_{M}\!\tr\Bigl(
\c^{2,0}_-\!\wedge * \bar\c^{0,2}_-\Bigr)
+\Fr{\bs_+\bbs_+}{6\pi^2}\int_M\!\tr\Bigl(\eta_-\!*\bar\eta_-\Bigr)
\cr
&
+\Fr{i\bs_+}{4\pi^2}\int_{M} \tr \Bigr(\c^{2,0}_-\wedge * F^{0,2}\Bigr) 
+\Fr{i\bbs_+}{4\pi^2}\int_M\tr\Bigl(\bar\c^{0,2}_-\wedge * F^{2,0}\Bigr).
}
}

This action functional indeed gives the desired  
equations for the fermionic zero-modes. After expanding the action 
functional $S$ we have the following terms relevant for fermionic 
zero-modes, 
\eqn\pld{
\eqalign{
S =-\Fr{1}{6\pi^2} \int_M\tr&\biggl(
i\bar\eta_-*\Dpp^*\p^{0,1}_+
+{i}\eta_-*\Dp^*\bar\p^{1,0}_+
+\Fr{3}{2}\chi^{2,0}_-\wedge * \Dpp\p^{0,1}_+
+\Fr{3}{2}\bar\chi^{0,2}_-\wedge *\Dp\bar\p^{1,0}_+
\cr
&
+\Fr{3i}{2}\chi^{2,0}_-\wedge * \Dpp^*\bar\l^{0,3}_+
+\Fr{3i}{2}\bar\chi^{0,2}_-\wedge *\Dp^*\l^{3,0}_+
\biggr) + \cdots.
}
}
From this we obtain the following fermionic equations of motion,
\eqn\ple{
\eqalign{
\Dpp^*\p^{0,1}_+=0,\cr
i\Dpp\eta_- +\Fr{3}{2}\Dpp^*\bar\chi^{0,2}_-=0,\cr
\Dpp\p^{0,1}_+ +i \Dpp^*\bar\l^{0,3}_+=0,\cr
\Dpp\bar\chi^{0,2}_-=0.\cr
}
}
We will see below that these give rise to exactly the 
required equations \bdg\ and  \kgb.

\subsec{The Path Integral}

The path integral of our model is localized to the
locus of the following equations, modulo $\CG$ symmetry, see \arj\ and
\ark,
\eqn\plf{
\eqalign{
\Dpp^*C^{0,3}=0,\cr
F^{0,2}=0,\cr
iF\wedge\o\wedge\o + \Fr{i}{2}[C^{3,0},C^{0,3}]
-\Fr{\zeta}{3}\o^3 I_E=0,\cr
}
}
and
\eqn\plg{
\eqalign{
\Da \phi_{++}=0,\cr
[\phi_{++}, C^{0,3}]=0,\cr
[\phi_{++},\phi_{--}]=0.
}
}
We call the moduli space defined by the eq.~\plf\ the
{\it extended} moduli space $\EM$ of EH connections (with
factor $\zeta$) or stable bundles.

Since the path integral is localized to integrable
connections $\Dpp^2=0$, the fermionic equations of motion in \ple\
become
\eqn\plh{
\eqalign{
\Dpp\eta_- =0,\cr
}\qquad
\eqalign{
\Dpp^*\p^{0,1}_+=0,\cr
\Dpp\p^{0,1}_+ =0,\cr
}\qquad
\eqalign{
\Dpp^*\bar\chi^{0,2}_-=0,\cr
\Dpp\bar\chi^{0,2}_-=0,\cr
}\qquad 
\Dpp^*\bar\l^{0,3}_+=0.
}
Thus the zero-modes of fermions
\eqn\pchfds{
\bar\eta_-,\,\p^{0,1}_+,\,\bar\chi^{0,2}_-,\,\bar\l_+^{0,3}
}
are elements of the cohomology group $\BH^{0,p}$
of the following Dolbeault complex \kah, 
\eqn\pli{
0\longrightarrow
\BC^{0,0}\, \ato{\Dpp}\,
\BC^{0,1}\, \ato{\Dpp}\,
\BC^{0,2}\, \ato{\Dpp}\,
\BC^{0,3}
\longrightarrow 0,
}
where $\BC^{0,\ell}:=\O^{0,\ell}(M,\End(E))$.
It is also easy to check that the above is isomorphic to the
versal deformation complex of the extended moduli space $\EM$ of stable
bundles.
Thus minus the index of the
above Dolbeault cohomology group corresponds to the net ghost number
violations in the path integral measure due to the zero-modes of fermions in
\pchfds. We have
\eqn\plj{
\eqalign{
\triangle&= -\#(\bar\eta_-)_0 +\#(\p^{0,1}_+)_0 
-\#(\bar\c^{0,2}_-)_0 + \#(\bar\l^{0,3}_+)_0
\cr
&=\sum_{q=0}^3 (-1)^{q+1} \dim \BH^{0,q}.
}
}
The net ghost number violation of the path integral
due to all the fermions -- the fermions in \pchfds\ and their conjugates -- 
is $(\triangle,\triangle)$.
The above index can be computed by applying the standard Riemann-Roch formula. 
We find 
\eqn\plk{
\triangle =
\int_M \opname{c}_1(M)\wedge\left( r \opname{c}_2(E) 
-\Fr{r-1}{2}\opname{c}_1(E)^2\right) 
-r^2(1 - h^{0,1}+h^{0,2}-h^{0,3}),
}
where $h^{p,q}$ denote the Hodge numbers of $M$.
We also note that a Hermitian vector bundle $E$ admits
an EH connection only if
\eqn\lubke{
\int_M \o\wedge\left( r \opname{c}_2(E) -\Fr{r-1}{2}\opname{c}_1(E)^2\right) 
\geq 0
}
and the equality holds if and only if $E$ is projectively
flat. 

Now we take a closer look at the path integral.
We note that the zero-modes of $\p^{0,1}_+$ and $\bar\l_+^{0,3}$,
thus $\BH^{0,1}$ and $\BH^{0,3}$,
correspond to local deformations of the extended moduli
space $\EM$. The other fermionic zero-modes  
$\bar\eta_-\in \BH^{0,0}$ and $\bar\chi^{0,2}_-\in\BH^{0,2}$  
will cause some trouble. Note that we have a decomposition
into trace and trace-free parts
\eqn\pll{
\BH^{0,q} = H^{0,q}(M) \oplus \widetilde{\BH}^{0,q}.
}
We call $\widetilde\triangle=\triangle -(-1+h^{0,1}-h^{0,2}+h^{0,3})$ 
the complex formal dimension
of $\EM$.
If we assume a situation that $\CG$ acts freely on the locus of
solutions of \plf, i.e., the connection is irreducible, 
the extend moduli space $\EM$
is an analytic space with the K\"{a}hler structure induced
from the $\CG$-equivariant K\"{a}hler form \plb.
The moduli space will not have the right complex dimension 
$\widetilde\triangle$
unless $\widetilde{\BH}^{0,2}=0$ as well.
In the ideal  situation  $\widetilde{\BH}^{0,0}=\widetilde{\BH}^{0,2}=0$,
the extended moduli space $\EM$ is smooth
and the zero-modes of $\p^{0,1}_+, \l_+^{3,0}$ span the holomorphic
tangent space.\foot{We will establish this later. We remark
that the case with $\BH^{0,3}\neq 0$ causes no problem
as this is associated with deformations of $\EM\supset \CM_{EH}$
along the direction of $C^{0,3}$. It would be a problem if we work
only with $\CM_{EH}$.} 
Thus the formal complex dimension 
is the actual dimension. 

However the assumption made above, in particular $\widetilde{\BH}^{0,2}=0$, 
is too naive.  We note that the obstruction to deformation of the
extended moduli space $\EM$ lies in $\widetilde{\BH}^{0,2}$.
In two complex dimensions Donaldson proved that one can
always achieve $\widetilde{\BH}^{0,2}=0$ after a suitable perturbation
of the metric. In three complex dimensions one can hardly expect
such a result to continue to hold. The assumption $\widetilde{\BH}^{0,0}=0$
is valid for a bundle $E$ with degree and rank coprime.

Let us see how the path integral deals with the above problems.
We assume, for simplicity, that our gauge group is $\SU(r)$,
so that $\End(E)$ is always trace-free (so we also should replace $\BH$ 
by $\widetilde{\BH}$).
Then the formal complex dimension $\triangle$ in \plj\ is given
by
\eqn\gshew{
\triangle = r\int_M \opname{c}_1(M)\wedge \opname{c}_2(E) -
(r^2-1)(1 - h^{0,1}+h^{0,2}-h^{0,3}),
}
instead of \plk. A typical observable of the theory is
the total $\CG$-equivariant K\"{a}hler form, after parity
change, $\widehat\varpi^\CG_T$ is given by \plb.
First we consider an idealistic case that 
$\BH^{0,0}=\BH^{0,2}=0$.
Then the correlation function $\langle \exp \widehat\varpi^\CG_T \rangle$
can be identified with the symplectic 
volume of $\EM$,
\eqn\plm{
\bigl\langle \exp \widehat\varpi_T^\CG \bigr\rangle
= \int_{\EM} \exp \widetilde\varpi_T=\vol(\EM).
}
If there are zero-modes for the anti-ghosts $\bar\chi^{0,2}_-$,
i.e. $\BH^{0,2}\neq 0$, the above correlation function is modified, 
\eqn\pln{
\bigl\langle \exp \widehat\varpi_T^\CG \bigr\rangle 
= \int_{\EM} \opname{e}(\CV)\wedge\exp \widetilde\varpi_T,
}
where $\opname{e}(\CV)$ denotes the Euler class of the anti-ghost
bundle $\CV$. One may consider correlation functions of
other observables $\widehat \CO^{r,s}$ with ghost
numbers $(r,s)$ given by
$\bs_+$ and $\bbs_+$ closed $\CG$ equivariant
differential forms $\CO^{r,s}$.
We have
\eqn\plo{
\left\langle \prod_{i=1}^\ell \widehat \CO^{r_i,s_i}\right\rangle
= \int_\EM \opname{e}(\CV)\wedge \widetilde \CO^{r_1,s_1}\wedge
\ldots \wedge \widetilde \CO^{r_\ell,s_\ell}
}
where $\widetilde \CO^{r,s}$ denotes the equivariant
differential form $\CO^{r,s}$ after the restriction and
reduction to $\EM$. The above correlation
function can be non-vanishing if 
\eqn\plp{
\sum_{i=1}^\ell (r_i, s_i) =(\triangle,\triangle),
}
due to the ghost number anomaly.
What is remarkable is that 
the path integral is well-defined even if the moduli
space $\EM$ does not satisfy conditions of unobstructedness 
like $\BH^{0,2}=0$. 

To understand this important point in more detail, let us 
look up some details about how the Euler class of
the anti-ghost bundle emerges.
The action function $S$ \tfaxk\ contains the following
Yukawa coupling involving the anti-ghost,
\eqn\angoa{
S = \Fr{i}{4\pi^2}\int\tr\left( \chi^{2,0}_-\wedge *[\phi_{++}, 
\bar\chi^{0,2}_-]\right)
+\cdots.
}
It also contains the following terms, solely from the
first line of the expression \tfaxk, depending on $\phi_{--}$,
\eqn\angob{
\eqalign{
S =- \Fr{1}{6\pi^2}\int\tr\biggl[
\phi_{--}\biggl(&
*\Da^*\Da \phi_{++}
-\Fr{1}{4}\bigl[C^{3,0},[\phi_{++},C^{0,3}]\bigr]\cr
&
-*\L [\p^{0,1}_+,\bar\p^{1,0}]
-\Fr{1}{4}[\l^{3,0}_+,\bar\l^{0,3}_+] 
\biggr)\biggr]
+\cdots.
}
}
Assuming, for simplicity, 
that there $\bar\eta_-$ has no zero-modes $(\BH^{0,0}=0$)
one can evaluate the correlation functions 
by solving the $\phi_{--}$ equations of motion
and replacing all the other fields, including $\phi_{++}$, 
by their zero-modes. Then
the only non-vanishing term in the action functional $S$
in the $\bs_+$ and $\bbs_+$ invariant neighborhood $\CC$ of the
fixed point locus comes from the expression \angoa, which can
be written as
\eqn\angoc{
S|_{\CC} = - \CF_{\a\bar\b i\barj}
\tilde\p^i_+ \tilde \p^{\barj}_+\tilde\c^\a_-\bar{\tilde\c}^{\bar \b}_-, 
}
where $\tilde\p^i_+$ and $\tilde\c^{\a}_-$ denote the zero-modes of
$(\p^{0,1}_+, \l^{3,0}_+)$ and $\chi^{2,0}_-$, respectively, 
and similarly for the conjugate fields. In the above
the indices $i$ and $\a$ run over $i=1,\ldots, \mathbf{h}^{0,1}+\mathbf{h}^{0,3}$ 
and $\a = 1,\ldots, \mathbf{h}^{0,2}$,
where $\mathbf{h}^{0,*} = \opname{\dim_\mathbb{C}} \BH^{0,*}$.
The expression 
$\CF^{\bar\a}{}_{\bar\b i\barj}\tilde\p^i_+\tilde\p^{\barj}_+$ denotes
the curvature two form of the anti-ghost bundle $\CV$ over $\EM$ --
the space of the zero-modes $a^i$ of $A^{0,1}$ and $C^{3,0}$ modulo $\CG$.
Consequently the expectation value, for example  
 $\langle \exp \widehat\varpi^\CG_T \rangle$,
becomes\foot{The determinant of the metric comes from integrating out the 
auxiliary fields.}
\eqn\angod{
\eqalign{
\langle \exp \widehat\varpi^\CG_T \rangle = \int_{\EM}&
\prod_{\ell=1}^{\triangle + \mathbf{h}^{0,2}} 
d a^\ell d a^{\bar \ell} d\p^\ell_+ d\p^{\bar \ell}
\prod_{\g=1}^{\mathbf{h}^{0,2}} d\c^\g_- d\c^{\bar\g}_-
\,\Bigl(\det h_{\a\bar\b}(a^\ell,a^{\bar\ell})\Bigr)\inv
\cr
&\times \exp\left( \CF_{\a\bar\b i\barj}(a^\ell,a^{\bar\ell})
\p^i_+ \p^{\barj}_+\c^\a_-\c^{\bar \b}_-  
+ \widetilde\o_{i\barj}(a^\ell,a^{\bar\ell})\p^i_+\p^{\barj}
\right),
}
}
which leads exactly to \pln.

\newsec{Deformations Of The Model}

In this section we study certain deformations of our $N_c=(2,0)$ model. 
The main purpose will be to be able to handle situations where 
non-stable bundles can occur, that is $\BH^{0,0}\neq0$ in the language 
of the last section, so there may be zero-modes for $\bar\eta_-$. 
In the moduli space of the theory considered until now, this situation 
introduces singularities. In our equivariant approach, which we 
implemented from the start, these type of singularities are however 
easily handled. The second deformation we consider might help to 
relate our extended model again to the unextended model based on 
EH bundles.

\subsec{Deformation To A ``Holomorphic'' $N_c=(2,0)$ Model}

In this subsection we consider a deformation of 
the original $N_c=(2,0)$ model. The resulting deformed 
model will have much better behavior than the original 
model when the effective target space $\CM_\zeta$
has singularities. This kind of deformation is originally due to 
Witten \tdYM\ and applied in a similar situation to the present
case in \Park. 
We will find the deformed and the original model 
as two special limits of a one-parameter family of models. 
In comparison with the discussion in \tdYM, we added the extra 
localization Fermi multiplets $\chi_-$; they will however be 
purely spectators, and the specialization to the K\"ahler 
case will simplify the procedure.

The original action was given by \tzaction. 
We saw that the path integral of the $N_c=(2,0)$
model is localized to the symplectic quotient 
$\CM_\zeta= (\CX\cap \m^{-1}(\zeta))/\CG$
of $\CX$ by $\CG$. 
Now we consider the following one-parameter family 
of $N_c=(2,0)$ theories, given by the action functional 
\eqn\defact{
\eqalign{
S(\z)_\l 
 =& S(\z)+\Fr{\l}{2}\bs_+\bbs_+\bigl\langle\phi_{--},\phi_{--}\bigr\rangle\cr
 =&-\bs_+\bbs_+\biggl(\bigl\langle\phi_{--}, \m-\z  
-\Fr{ \l}{2}\phi_{--}\bigr\rangle
-\bigl\langle\eta_-,\bar\eta_-\bigr\rangle
+\bigl\langle h_{\a\bar \b}\c^\a_-,\c^{\bar \a}_- \bigr\rangle\biggr)\cr
& 
+i\bs_+\bigl\langle\c^\a_-,\ES_\a\bigr\rangle 
+\bbs_+\bigl\langle\c^{\bar \a}_-, \ES_{\bar \a}\bigr\rangle.
}
}
If we set $\l=0$ we retain the original action. 
Since the $\l$-dependent term by which we deform is $\bs_+$ and $\bbs_+$ 
closed, the theory does not depend on $\l$, as long as $\l\neq0$. 
The models with $\l=0$, and $\l\neq0$ can be different since new fixed points 
can flow in from infinity in the field space \tdYM.  
For $\l\neq0$ the path integral localixes to 
the critical points of $I=\langle\m-\z,\m-\z\rangle$, while the original 
theory, at $\l=0$, is localized to the zeros (trivial critical points) 
of $I$. 

Next, we add local $\bs_+$ and $\bbs_+$ closed observables
$-\widehat\varpi^\CG$, $i\langle\phi_{++},\z\rangle$ and 
$-\Fr{\e}{2}\langle\phi_{++},\phi_{++}\rangle$ 
to this action functional, basically for regularization. 
We get 
\eqn\hdefact{
\eqalign{
S_h(\z,\e)_\l 
 =& S(\z)_\l -i\bs_+\bbs_+ \CK +i\langle\phi_{++},\z\rangle
 +\Fr{\e}{2}\langle\phi_{++},\phi_{++}\rangle\cr
 =& -i\langle\phi_{++},\m-\z\rangle -\widehat\varpi (\p_+,\bar\p_+)
+\Fr{\e}{2}\langle\phi_{++},\phi_{++}\rangle \cr
&-\bs_+\bbs_+\biggl(\bigl\langle\phi_{--}, \m-\z 
-\Fr{ \l}{2}\phi_{--}\bigr\rangle
-\bigl\langle\eta_-,\bar\eta_-\bigr\rangle
+\bigl\langle h_{\a\bar \b}\c^\a_-,\c^{\bar \a}_- \bigr\rangle\biggr)\cr
& +i\bs_+\!\bigl\langle\c^\a_-,\ES_\a\bigr\rangle 
+\bbs_+\!\bigl\langle\c^{\bar \a}_-, \ES_{\bar \a}\bigr\rangle.
}
}
The partition sum of this model computes the expectation value 
$\bigl\langle e^{-\CO(\z,\e)}\bigr\rangle$ in the deformed theory 
\defact, where $\CO(\z,\e)$ denotes the extra contributions in the 
action above. 

For $\l\neq0$, we can integrate out the $N_c=(2,0)$ gauge multiplet 
$(\phi_{--},\eta_-,\bar\eta_-,D)$. We are then left with 
\eqn\tzgac{
\eqalign{
S_h(\z,\e)_\l 
 =&
-i\left\langle\phi_{++}, \m-\z \right\rangle 
- \widehat\varpi ( \p_+,\bar\p_+)
+\Fr{\e}{2}\left\langle\phi_{++},\phi_{++}\right\rangle
\cr
&-\bs_+\bbs_+\bigl\langle h_{\a\bar \b}\c^\a_-,\c^{\bar \a}_- \bigr\rangle
 +i\bs_+\!\bigl\langle\c^\a_-,\ES_\a\bigr\rangle 
+\bbs_+\!\bigl\langle\c^{\bar \a}_-, \ES_{\bar \a}\bigr\rangle.\cr
&+\Fr{1}{2\l}\bs_+\bbs_+\left\langle\m-\z  ,\m-\z\right\rangle
+\CO(1/\l^2).
}
}
If we take the limit $\l\rightarrow 0$, while $\l\neq 0$,
we see that the dominant contributions to the
path integral come from the critical points of
$I=\langle\m-\zeta,\m -\zeta\rangle$. Note that this includes 
the trivial critical points $\m=\z$, whose contributions give 
the path integral of the original model, defined by \tzaction, 
with the insertions of the observables added above. However, 
in general we also get contributions from higher critical points. 
So we do not get back the original model. The contributions of 
the higher critical points, for which $I\neq0$, are proportional 
to $e^{-I/2\e}$, for $\e\to0$ (this can be seen by integrating 
out $\phi_{++}$). Therefore, we can easily extract the 
contribution from the original model. On the other hand, as the 
theory is independent of $\l\neq0$, this limit is the same as the 
theory for \emph{any} value $\l\neq0$. 

Now consider the limit $\l\to\infty$, to remove all the 
$\l$-dependent terms from \tzgac. 
We call this model a holomorphic $N_c=(2,0)$ model.\foot{
This name is inspired by the holomorphic Yang-Mills theory \Park, 
which arises from Donaldson-Witten theory in this way.} 
The path integral of this theory is localized to critical 
points of $I=\langle\m-\z,\m-\z\rangle$, which shows that 
indeed this limit is the same as the deformed model given by 
\tzgac\ for finite $\l$.

\subsec{A Use Of $S^1$ Symmetry}

The extended equations \plf\ we have 
may be very useful. On the extended moduli space $\EM$
of EH connections we have the natural $S^1$-action
\eqn\pca{
S^1: C^{0,3}\rightarrow e^{i\th} C^{0,3},
}
which preserves the complex and the K\"{a}hler structure.
Thus any cohomological computation can be
further localized to the fixed point locus of this $S^1$-action.
For the $\SU(2)$ case we are concentrating on it is easy
to determine the fixed points. We have two branches.

\begin{itemize}

\item Branch (i)

$\phi_{++}=0$ and the $\SU(2)$ symmetry is unbroken.
Then we have a trivial fixed point where simply $C^{0,3}=0$ 
where we have EH connections.

\item Branch (ii)

$\phi_{++}$ is a constant diagonal trace-free matrix.
The nontrivial fixed points occur if the gauge symmetry
can undo the $S^1$-action. For this the $\SU(2)$ symmetry
should be broken to $\U(1)$. Thus the gauge bundle splits, 
$E_A = L \oplus L^{-1}$ where $A\in \CA^{1,1}$. 
Furthermore $C^{0,3}$ and $C^{3,0}$
become
\eqn\vupb{
C^{0,3}=\left(\matrix{ 0 & \g\cr 0 & 0}\right),\qquad
C^{3,0}=\left(\matrix{ 0 & 0\cr \bar\g & 0}\right),
}
where $\g$ is a section of $K^{-1}\otimes L^2$, with
$K$ denoting the canonical line bundle of our K\"{a}hler $3$-fold.
Then we have the following fixed point equations
\eqn\pcb{
\eqalign{
F^{0,2}_L&=0,\cr
iF_L\wedge\o\wedge\o -\Fr{1}{2}\g\wedge\bar\g&=0.
}\qquad
\rd^*_L \g =0,
}
where $F_L$ denotes the curvature of the line bundle $L$.
Obviously we have a nontrivial solution if $\deg(L) > 0$.
If $\g=0$ we can have abelian EH connections, and also if 
$\deg(L)=0$.

\end{itemize}

The equations \pcb\ are analogous to the abelian 
Seiberg-Vafa-Witten equations \SWinv\VW; they may be equally 
powerful. Thus we expect that the above equations may contain
all the nontrivial  information about the Donaldson-Witten type
theory on K\"{a}hler $3$-folds. It should be possible
to establish our conjecture quite rigorously. 
Here we will only sketch the idea.

As a first step we map the $N_c=(2,0)$ model
defined by the action functional $S$ \tfaxk\ to its deformed
version, following the discussions in Sect.~5.1.
The action functional is then defined by
\eqn\tvaxx{
\eqalign{
S_h(\e) = &\Fr{1}{4\pi^2}
\bs_+\bbs_+\int_{M}\tr\left(\c^{2,0}_-\wedge * \bar\c^{0,2}_-\right)
\cr
&
+\Fr{i}{4\pi^2}\bs_+\int_{M} \tr \biggr(\c^{2,0}_-\wedge * F^{0,2}\biggr)
+\Fr{i}{4\pi^2}
 \bbs_+\int_M\tr\biggl(\bar\c^{0,2}_-\wedge * F^{2,0}\biggr)
\cr
&
-i\bs_+\bbs_+\CK_T
+\Fr{\e}{4\pi^2}\int_M \Fr{\o^3}{3!}\tr(\phi_{++}^2),
}
}
where $\CK_T$ is given by \pla.
As we established earlier the partition function of this theory 
for $\e=0$ is the correlation function \pln\ with the same conditions.
If the reducible connections are unavoidable we turn on
$\e$ to regularize and utilize the non-abelian localization.

Examining the supersymmetry transformation laws of the holomorphic 
$C^{3,0}$ and the Fermi $\c^{2,0}$ multiplets, we can see 
that the $S^1$-action \pca\ should be extended as follows
\eqn\pcaex{
\eqalign{
S^1:(C^{0,3}, \bar\l^{0,3}_+, \bar\chi^{0,2}_-, H^{0,2})
\rightarrow \xi (C^{0,3}, \bar\l^{0,3}_+, \bar\chi^{0,2}_-, H^{0,2}),\cr
S^1:(C^{3,0}, \l^{3,0}_+, \chi^{2,0}_-, H^{2,0})
\rightarrow \bar\xi (C^{3,0}, \l^{3,0}_+, \chi^{2,0}_-, H^{2,0}),\cr
}
}
where $\xi\bar\xi=1$. Thus the above fields are now charged
under $S^1$. A problem might be that the above $S^1$-action
is not a symmetry of the action functional.\foot{
This is due to a term like $\tr(\c^{2,0}_-\wedge *\Dpp\bar\p^{0,1}_+)$.
}
However the $S^1$-action  preserves the supersymmetry
transformation laws as well as the localization equations.
Thus we can use it anyway. Now we modify the transformation
laws of the charged fields under the $S^1$ by extending
the $\CG$-equivariant cohomology to $\CG\times S^1$;
\eqn\familiar{
\bs^2_+=0,\qquad \{\bs_+,\bbs_+\} = -i\phi^a_{++}\CL_a -i m
\CL_{S^1},\qquad
\bbs_+^2=0.
}

We use the same form of the deformed action functional \tvaxx\
but with the new transformation laws for supercharges according
to \familiar.
We obtain a new $N_c=(2,0)$ supersymmetric action 
functional\foot{We put $\e$ to zero. We can turn on $\e$ whenever necessary.}
\eqn\ultimate{
\eqalign{
S_h(m,\e) = 
&\Fr{1}{4\pi^2}
\bs_+\bbs_+\int_{M}\tr\left(\c^{2,0}_-\wedge * \bar\c^{0,2}_-\right)
\cr
&
+\Fr{i}{4\pi^2}\bs_+\int_{M} \tr \biggr(\c^{2,0}_-\wedge * F^{0,2}\biggr)
+\Fr{i}{4\pi^2}
 \bbs_+\int_M\tr\biggl(\bar\c^{0,2}_-\wedge * F^{2,0}\biggr)
\cr
&
 -\widehat\varpi^\CG_T
-im  H_{S^1},
}
}
where $H_{S^1}$ is the bosonic Hamiltonian of the $S^1$-action,
\eqn\hecds{
H_{S^1} =\Fr{1}{24\pi^2} \int_M \tr \left(C^{3,0}\wedge C^{0,3}\right).
}
The first and second lines in the action functional 
localize the path integral to the locus $\Dpp^*C^{0,3}=F^{0,2}=0$. 
The first term in the third line 
further localize the path integral to the locus
$\m_T=0$. 
For simplicity we assume that there are no
zero-modes of $\chi^{2,0}_-$.
Then the partition function of the model reduces
to\foot{We remark that the
action functional contains the mass term for the anti-ghosts 
$\chi^{2,0}_-$ and $\bar\chi^{0,2}_+$. If there are no zero modes 
for anti-ghost such the term plays no roles.
If there are zero-modes of anti-ghosts we have to include
contribution from the anti-ghost bundles and the mass term.
Then the partition function $Z$ becomes
$$
Z = \int_{\EM} det(\CF -i m I)\exp\left(im H_{S^1} +\widetilde \varpi_T\right),
$$
where $\CF -i m I$ is the $S^1$-equivariant
curvature two form of the  anti-ghost bundle $\CV$ over $\EM$.
}
\eqn\edfd{
Z = \int_{\EM} e^{im H_{S^1} +\widetilde \varpi_T},
}
where 
$\widetilde \varpi_T$ is the K\"{a}hler form of $\EM$,
obtained by the restriction and reduction from our equivariant
K\"{a}hler form $\widehat\varpi^\CG_T$ \plb.
Thus the partition function is given by the familiar DH integral
formula over a finite dimensional K\"{a}hler manifold $\EM$ \DH\Wu.
It is therefore an integral over the set of critical points
of $H_{S^1}$, which is the same as the fixed point locus
of the $S^1$-action on $\EM$.
Thus we have the same two branches.

The following is a formal argument since we do not understand
the compactification of $\EM$. However it will be sufficient
to serve our purpose.
We will just apply the
exactness of the stationary phase integral.
By setting $m\rightarrow \infty$ we may have

\begin{itemize}

\item Branch (i)

Note that the value of the Hamiltonian $H_{S^1}$ is zero
at Branch (i). So its contribution to the integral is
simply the volume of $\CM_{EH}$ weighted by
the one loop determinant of due to the normal bundle
$N(\CM_{EH})$ in $\EM$. 
Note that such one loop determinant contains weight
$m^{-s}$ where $s$ denotes codimension
of $\CM_{EH})$ in $\EM$.
Thus
\eqn\specqa{
Z(i) \sim \Fr{1}{m^s}\vol(\CM_{EH})\times \cdots.
}
The unwritten part is due to contribution from the normal bundle
$N(\CM_{EH})$,
while we extracted its dependence on $m$.

\item Branch (ii)

Note that
the value of the Hamiltonian at Branch (ii) is
$$
H_{S^1}=\Fr{1}{12\pi}\deg(L) := \Fr{1}{24\pi^2}
\int \opname{c}_1(L)\wedge\o\wedge\o,
$$
where $L$ is a line bundle defined in \pcb.
Thus
\eqn\specqb{
Z(ii) \sim \sum_{L} \Fr{1}{m^{s^\pr}}\int_{\CF(L)}
\exp\left( -\Fr{im}{12\pi}\deg(L)
+\widetilde\o\mid_{\CF(L)}\right)\times \cdots
}
where $\CF(L)$ denotes the fixed point locus, $s^\pr$ denotes
its codimension and $ \widetilde\o|_{\CF(L)}$ denote the K\"{a}hler form
on $\CF(L)$. 
The unwritten part is due to contributions from the normal bundle 
over the fixed point locus,
while we extracted its dependence on $m$.

\end{itemize}

We assume that $s < s^\pr$, otherwise the above formal
formula does not make sense. Then one can take $m=0$.
Since the original formula was smooth in the limit
of the reduction to the symplectic volume of $\EM$
the poles in $Z(i)$ and $Z(ii)$ should cancel order by
order. Thus we have
\eqn\onald{
\eqalign{
\vol(\CM_{EH})\sim \sum_{L} &\Fr{1}{(s^\pr-s)!}
 \left(\Fr{im}{12\pi}\deg(L)\right)^{s^\pr -s}
\cr
&
\times
\int_{\CF(L)}
\exp\left( -\Fr{im}{12\pi}\deg(L)
+\widetilde\varpi|_{\CF(L)}\right)\times \cdots
}}
and
\eqn\oanald{
\eqalign{
\vol(\EM)\sim \sum_{L} &\Fr{1}{s^\pr!}
 \left(\Fr{im}{12\pi}\deg(L)\right)^{s^\pr}
\cr
&
\times
\int_{\CF(L)}
\exp\left( -\Fr{im}{12\pi}\deg(L)
+\widetilde\varpi|_{\CF(L)}\right)\times \cdots.
}}

We conclude that the above formal evaluation gives evidence for 
our conjecture that Seiberg-Vafa-Witten type invariants
defined by the equation \pcb\ should be equivalent to
the Donaldson-Witten type invariants on a K\"{a}hler $3$-fold.
It is possible to perform a similar analysis for the case with
anti-ghost zero-modes, which makes life more complicated
but does not alter the essential points advocated above.

\newsec{Specialized Models}

We will now shortly comment on properties of the model in some 
special situations when the K\"ahler 3-fold has additional 
symmetries, that is more reduced holonomy.

\subsec{Reduction To A K\"{a}hler Surface}

In this subsection we perform a dimensional reduction
of our models on a K\"{a}hler $3$-fold $M$ to a complex
K\"{a}hler surface $M_2$. We first assume that $M$ is a product
manifold $M_3= M_2\times\mathbb{C}$ and then remove dependence
of our fields on $\mathbb{C}$. We have the following correspondence
\eqn\era{
\eqalign{
&A^{0,1}\rightarrow A^{0,1},\, \s,\cr
&\p^{0,1}_+\rightarrow \p^{0,1}_+,\, \bar\eta_+,\cr
&\chi^{2,0}_-\rightarrow \p^{1,0}_-,\, \chi^{2,0}_-,\cr
&H^{0,2}\rightarrow H^{0,1},\, H^{0,2},\cr
&C^{0,3}\rightarrow B^{0,2},
}
}
as well as the corresponding decomposition for their Hermitian conjugates.
The other fields $(\phi_{\pm\pm},\eta_-,\bar\eta_-,D)$ remain
as they were.
Thus we obtain a $N_c=(2,2)$ model.
Similarly the equation \plf\ for  the extended EH connection
reduces to the Vafa-Witten equations. Furthermore our equation \pcb\ for
branch (ii) fixed point become the Abelian Seiberg-Witten equations.
Thus our conjecture on Donaldson-Witten type invariants on a K\"{a}hler
$3$-fold becomes a fact \SWinv. The model we obtain is exactly
the Vafa-Witten theory of a twisted $\CN=4$ super-Yang-Mills theory
on the K\"{a}hler surface \VW\DPS\LaL.

Now instead of the above trivial reduction we consider a product
manifold $M=M_2\times \S$, where $\S$ is a 2-torus.
Then we can follow the same steps with the same sort of assumption
as \BJSV\ to conclude that the models discussed in the previous subsection
are equivalent to the topological sigma model of Vafa and Witten \VW.
Thus the stringy Donaldson-Witten invariants on a K\"{a}hler surface
may be obtained from formulas like \onald\ and \oanald\ 
on the product $3$-fold. This supports an earlier suspicion of one of 
the authors that a stringy generalization of Donaldson-Witten theory 
as discussed in \DPS\ does not give information beyond Seiberg-Witten, 
since the Seiberg-Vafa-Witten type invariants on a manifold $M_2\times\S$
most likely are just the Seiberg-Witten invariants on $M_2$.

\subsec{The Model On Calabi-Yau 3-Folds}

We now shortly comment on the case that the K\"{a}hler $3$-fold
$M$ is  Calabi-Yau with  holomorphic $3$-form $\o^{0,3}$. 
For the Calabi-Yau case the $N_c=(2,0)$ supersymmetry enhances 
to $N_c=(2,2)$ supersymmetry. We will come back to this model in 
more detail in a forthcoming paper \HPE.

We argued in \HPB\ that our model is the world-volume theory
of parallel type IIB (Euclidean) $D5$-branes wrapped on
the $CY_3$. We show that the $\CG$-equivariant degrees of 
freedom correspond to the bulk degrees of freedom
transverse to the (Euclidean) $D5$-branes. We use such a 
correspondence as supporting evidence that our path
integral should be well-defined in any situation.

We consider the $N_c=(2,0)$ theory with supercharges
$\bs_+$ and $\bbs_+$ defined in the previous
section specializing to a Calabi-Yau $3$-fold $M$ with
a holomorphic $3$-form $\o^{0,3}$.
Using the non-degeneracy
of $\o^{0,3}$ we may redefine the fields 
$(\bar\chi^{0,2}_-,H^{0,2},\l^{0,3}_+, C^{0,3})$ as
\eqn\pma{
\p^{0,1}_-,\, H^{0,1},\,\eta_+, \s,
}
where\foot{The anti-holomorphic Hodge star operator $\bar*$ is 
defined by $\bar*\a=*\bar\a$. Acting on
a $(p,q)$-form on a complex $d$-fold gives a $(d-p, d-q)$,
$$
\bar*:\O^{p,q}(M) \rightarrow \O^{d-p, d-q}(M).
$$
}
\eqn\pmb{
\eqalign{
\bar\chi^{0,2}_- &=  \bar*(\o^{3,0}\wedge  \p^{0,1}_-),\cr
H^{0,2} &=  \bar*(\o^{3,0}\wedge  H^{0,1}),\cr
}\qquad
\eqalign{
\l^{0,3}_+&= \eta_+\o^{0,3},\cr
C^{0,3}&= \s \o^{0,3}.
}
}

It is not difficult to show that the action functional 
$S$ has additional global supersymmetries generated
by $\bs_-$ and $\bbs_-$. We have the following
diagrams to be compared with \pxd;
\eqn\pmc{
\def\normalbaselines{\baselineskip20pt
\lineskip3pt \lineskiplimit3pt}
\matrix{
 \bar\s    & \mapr & \eta_+ & \mapl & \phi_{++}  \cr
 \mapd     &       & \mapd  &       & \mapd      \cr
\bar\eta_- & \mapr & D      & \mapl & \bar\eta_+ \cr
\mapu      &       & \mapu  &       & \mapu      \cr
\phi_{--}  & \mapr & \eta_- & \mapl & \s         \cr
}\quad,\qquad
\matrix{
\p^{0,1}_- & \mapl  & A^{0,1} & \mapr  & \p^{0,1}_+ \cr
           & \maprd &         & \mapld &            \cr
           &        & H^{0,1} &        &
}.
}
The four supercharges satisfy the following
anti-commutation relations
\eqn\pmd{
\eqalign{
\{\bs_\pm,\bs_\pm\}=0,\cr
\{\bbs_\pm,\bbs_\pm\}=0,\cr
}
\qquad
\eqalign{
\{\bs_+,\bbs_+\} &= -i\phi_{++}^a\CL_a,\cr
\{\bs_+,\bbs_-\}&=-i\s^a\CL_a,\cr 
\{\bs_-,\bbs_+\} &= -i\bar\s^a\CL_a,\cr
\{\bs_-,\bbs_-\} &= -i\phi^a_{--}\CL_a,\cr
}\qquad
\eqalign{
}
\qquad
\eqalign{
\{\bs_+,\bs_-\}=0,\cr
\{\bbs_+,\bbs_-\}=0.\cr
}
}
The above anti-commutation relations define a balanced $\CG$-equivariant 
Dolbeault 
cohomology on the space $\CA$ of all connections \DPS.
Thus our model becomes a $N_c=(2,2)$ model.

The action functional $S$ in \tfaxk\ can be rewritten in a form 
showing manifest $N_c=(2,2)$ symmetry,
\eqn\pme{
\eqalign{
S = \bs_+\bbs_+\bs_-\bbs_-\biggl(
\CK
-\Fr{1}{6\pi^2}\int_M \tr (\s *\bar\s)\biggr)
+\bs_+\bs_-\CW(A^{0,1})
+\bbs_+\bbs_-\overline \CW(A^{1,0}),
}
}
where $\CK$ is the K\"{a}hler potential on the space $\CA$ of all
connections,
\eqn\kpos{
\CK=\Fr{1}{24\pi^2}\int_M \k\tr\left( F\wedge F\right)\wedge \o,
}
and $\CW(A^{0,1})$ is the holomorphic Chern-Simons form,
\eqn\hcs{
\CW(A^{0,1})=\Fr{1}{8\pi^2}\int_M \o^{3,0}\wedge \tr \Bigl(A\wedge\bar\rd A
+\Fr{2}{3}A\wedge A\wedge A\Bigr).
}

We remark that the above action functional can be obtained by 
the dimensional
reduction of the $(1+1)$-dimensional $N_{ws}=(2,2)$ spacetime supersymmetric 
linear gauged sigma model in two real dimensions, whose
target space is the space $\CA$ of all connections on a Calabi-Yau
$3$-fold $M$ \HPB. In \HPB\ we interpreted the model as the
matrix string theory \Motl\BS\matrixstring\ compactified on a Calabi-Yau
by regarding $\CA$ as the configuration space of all D-branes
wrapped on the Calabi-Yau.

\newsec{Discussion And Conclusion}

In this paper we studied an extended moduli problem of stable 
bundles on K\"ahler 3-folds, using topological field theory. 
The partition function of the topological field theory gives 
a concrete formula to calculate natural 
generalizations of Donaldson-Witten type invariants for higher 
dimensional K\"ahler manifolds. The bare problem of stable 
bundles in 3 complex dimensions generically is obstructed, 
which is reflected in the infinite number of anti-ghost zero-modes 
in the corresponding model. We argued that in order to reduce 
this to a finite number, we had to extend the model by adding a 
$(3,0)$-form field. This extended moduli problem indeed 
gives rise to a finite number of zero-modes, and therefore 
also a finite dimensional moduli space. However, the model 
may still have anti-ghost zero-modes, which would make the 
moduli space non-smooth. However, the partition function 
and the correlation functions can still be well defined, 
by using the Euler class of the corresponding anti-ghost bundle.  

Another potential problem was the appearance of zero-modes for the 
ghosts, corresponding to the possible appearance of strictly 
semi-stable bundles. We saw that we could deform the model such that 
we are able to deal with thissituation. This deformation is similar 
to the one proposed in Donaldson theory in \tdYM. 

Stable bundles also appear as the BPS sector of string theory, 
interpreted as BPS configurations of D-branes wrapped around the 
K\"ahler manifold. It would be interesting to see if the extended 
moduli problem also has a string interpretation, though at firs
sight this does not seem the case, as we have no natural candidate 
for the additional $(3,0)$-form. 

The general mathematical cohomological problem has a generalization 
to higher dimensional K\"ahler manifolds. However, we could not 
implement these ideas into a topological field theory setting. 
The only solution could lie in the interpretation of the higher 
even forms as gauge parameters rather than field strengths 
(obstructions). However, we do not know anyway in which this 
could happen in the nonabelian case. It is interesting to compare to 
string theory, where there are strong hints towards ''nonabelian`` 
higher form gauge transformations.

\ack

We are grateful to Erik and Herman Verlinde for encouragement and 
discussions. CH was supported by the FOM foundation and DOE grant 
\# DE-FG02-96ER40559. JSP was supported by a pioneer fund of NWO and 
DOE grant \# DE-FG02-92ER40699.

\appendix

\newsec{Supersymmetry Transformation Laws}

In this Appendix we give the explicit $N_c=(2,0)$ supersymmetry 
transformation rules of the fields of our model discussed in Sect.~$4$. 
The transformation laws of the {\it gauge multiplet} 
$(\phi_{--},\eta_-,\bar\eta_-,D)$ and of $\phi_{++}$ are given by 
\eqn\pab{
\eqalign{
\bs_+ \phi_{--} = i\eta_-,\cr
\bbs_+\phi_{--}=i\bar\eta_-,\cr
}
\qquad
\eqalign{
\bs_+ \eta_-&=0,\cr
\bbs_+\eta_- &=+i D + \Fr{1}{2}[\phi_{++},\phi_{--}],\cr
\bs_+\bar\eta_-&=-iD +\Fr{1}{2}[\phi_{++},\phi_{--}],\cr
\bbs_+\bar\eta_-&=0,
}\qquad \eqalign{
\bs_+\phi_{++}=0,\cr
\bbs_+\phi_{++}=0.\cr
}
}
We had two sets of holomorphic multiplets and their 
anti-holomorphic partners.
The transformations follow those given in \are. 
One set of holomorphic multiplets is $(A^{0,1},\p^{0,1}_+)$  
with anti-holomorphic partners $(A^{1,0},\bar\p^{1,0}_+)$,
\eqn\baxa{
\eqalign{
\bs_+ A^{0,1} &=i\p^{0,1}_+,\cr
\bbs_+ A^{0,1}&=0,\cr
\bs_+ A^{1,0}&=0,\cr
\bbs_+ A^{1,0}&=i\bar\p^{1,0}_+,\cr
}\qquad
\eqalign{
\bs_+ \p^{0,1}_+&=0,\cr
\bbs_+\p^{0,1}_+ &=- \Dpp \phi_{++},\cr
\bs_+\bar\p^{1,0}_+&=-\Dp \phi_{++},\cr
\bbs_+\bar\p^{1,0}_+&=0.\cr
}
}
The other  holomorphic multiplet is $(C^{3,0},\l^{3,0}_+)$ 
with anti-holomorphic partner $(C^{0,3},\bar\l^{0,3}_+)$,
\eqn\baxb{
\eqalign{
\bs_+ C^{3,0} &=i\l^{3,0}_+,\cr
\bbs_+ C^{3,0}&=0,\cr
\bs_+ C^{0,3}&=0,\cr
\bbs_+ C^{0,3}&=i\bar\l^{0,3}_+,\cr
}\qquad
\eqalign{
\bs_+ \l^{3,0}_+&=0,\cr
\bbs_+\l^{3,0}_+ &=-i [\phi_{++}, C^{3,0}],\cr
\bs_+\bar\l^{0,3}_+&=-i[\phi_{++},C^{0,3}],\cr
\bbs_+\bar\l^{0,3}_+&=0.\cr
}
}
Finally we have Fermi multiplets $(\chi^{2,0}_-, H^{2,0})$
and anti-Fermi multiplets  $(\bar\chi^{0,2}_-, H^{0,2})$, 
with transformation rules as in \ferm, using the holomorphic 
section $\EJ$ given in \pke, we get 
\eqn\baxc{
\eqalign{
\bs_+\chi^{2,0}_- &= -H^{2,0},\cr
\bbs_+\chi^{2,0}_-&= -\Dp^* C^{3,0},\cr
\bs_+\bar\chi^{0,2}_- &= -\Dpp^*C^{0,3},\cr
\bbs_+\bar\chi^{0,2}_-&= -H^{0,2},\cr
}
\qquad
\eqalign{
\bs_+H^{2,0} &=0,\cr
\bbs_+ H^{2,0} &= -i [\phi_{++},\chi^{2,0}]
+i [*\p^{0,1}_+*, C^{3,0}]
+i \Dp^*\l^{3,0}_+
        , \cr
\bs_+ H^{0,2} &= -i [\phi_{++},\bar\chi^{0,2}]
+i [*\bar\p^{1,0}_+*, C^{0,3}]
+i \Dpp^*\bar\l^{0,3}_+
       , \cr
\bbs_+H^{0,2} &=0.
}
}

\newsec{Some Properties Of $\EM$}

This is a mathematical digression to establish a property
of the extended moduli space. 
First  we recall a theorem \Kobayashi\Kim\ on the moduli
space $\CM_{EH}$ of EH connections -- if 
$\widetilde{\mathbf\BH}^{0,0}=0$ the moduli space $\CM_{EH}$
is a complex analytic space. It is nonsingular at a neighborhood
of a connection if  $\widetilde{\mathbf\BH}^{0,2}=0$ 
and its tangent space is naturally isomorphic to the space
of $\BH^{0,1}$. Here $\widetilde{\mathbf\BH}^{0,*}$
denotes the cohomology group defined by tracefree endomorphisms.
We refer to \Kobayashi Ch.\ VII.3 for details on the notations.

Now we  state an analogous theorem about the extended moduli
space $\EM$ of EH connections on a complex K\"{a}hler
$3$-fold  - if $\widetilde{\mathbf\BH}^{0,0}=0$
the moduli space $\EM$
is a complex analytic space. It is nonsingular at a neighborhood
of an extended connection if  $\widetilde{\mathbf\BH}^{0,2}=0$ 
and its tangent space is naturally isomorphic to the space
$\BH^{0,1}\oplus \BH^{3,0}$.
The extended moduli space $\EM$
is a smooth K\"{a}hler manifold with the formal dimension
equal to the actual dimension if 
$\widetilde{\mathbf\BH}^{0,0}=\widetilde{\mathbf\BH}^{0,2}=0$.

The proof of the above theorem  is similar to that of the Einstein-Hermitian
case  \Kobayashi. 
Given an  extended EH connection $\bar\mathfrak{D}$,
a nearby deformation $\Dpp +\a$, $C^{3,0} + \b$ 
is governed by the equations
\eqn\kimf{
\eqalign{
\Dpp \a + \a\wedge\a =0,\cr
\Dpp^*\a=0,\cr
\L(\Dpp \b +\a\wedge\b)=0.
}
}
We only need to consider the last equation since the theorem
quoted above already dealt with the first two equations.
The last equation
has
the following orthogonal decomposition
\eqn\Kimg{
\Dpp \b + \a\wedge \b=0
\leftrightarrow
\left\{\eqalign{
&\Dpp\left(\b + \Dpp^*\circ G(\a\wedge\b)\right)=0,\cr
&\Dpp^*\oplus \Dpp\circ G(\a\wedge\b)=0,\cr
&H(\a\wedge\b)=0,
}\right.
}
where $G$ is Green's operator and $H$ is the harmonic projection.
We define Kuranishi map $k^\pr$ 
\eqn\Kimh{
k^\pr: \BC^{3,0}\rightarrow \BC^{3,0},
\qquad k^\pr(\b) = \b + \Dpp^*\circ G(\a\wedge\b).
}
Then, from the first equation on the right of \Kimg\ 
we have $\Dpp (k^\pr(\b))=0$, while $\Dpp^*(k^\pr(\b))=0$
by the dimensional reason. 
Thus we obtain $\L \Dpp (k^\pr(\b))=0\rightarrow 
\Dpp^* (k^\pr(\b))=0$. Consequently we have
\eqn\Kimi{
k^\pr(\b) \subset \BH^{3,0}.
}

Now we examine if the Kuranishi map is invertible for
a given $\r \in  \BH^{3,0}$, i.e., $\b =k^{\pr -1}(\r)$
and  $\L(\Dpp\b +\a\wedge\b)=0$. Taking the orthogonal
decomposition of $\a\wedge\b$
one finds that
\eqn\Kimj{
\L(\Dpp\b +\a\wedge\b)=\L \Dpp^*\circ\Dpp\circ G(\a\wedge\b)
+ \L(H(\a\wedge\b)).
}
Note that $\L(H(\a\wedge\b))$ is in $\widetilde{\mathbf\BH}^{2,0}$, 
which is isomorphic
to $\widetilde{\mathbf\BH}^{0,2}$. By our assumption 
we have $H(\a\wedge\b))=0$.
Denoting $\g = \Dpp\a + \a\wedge\a$ and $\d = \Dpp\b +\a\wedge\b$
we have
\eqn\Kimk{
\eqalign{
\d &= \Dpp^*\circ G(\Dpp\a \wedge\b - \a\wedge\Dpp\b)\cr
&=\Dpp^*\circ G(\g\wedge\b + \a\wedge\d)\cr
&=\Dpp^*\circ G(\a\wedge\d),
}
}
where we used the fact that $\g=0$ for $\widetilde{\mathbf\BH}^{0,2}=0$.
Applying the following estimate
\eqn\Kiml{
\|\Dpp^*\circ G v\|_{2,k+1}\leq c\|v\|_{2,k},
}
we have
\eqn\Kimn{
\eqalign{
\|\d\|_{2,k}&\leq \|\d\|_{2,k+1}=\|\Dpp^*\circ G(\a\wedge\d)\|_{2,k+1}
\cr
&\leq c\|\d\|_{2,k}\cdot \|\a\|_{2,k}.
}
}
Taking $\a$ sufficiently close to $0$ so that $\|\a\|_{2,k} <1/c$,
we conclude $\d=0$. Thus the Kuranishi map $k^\pr$
is invertible if $\widetilde{\mathbf\BH}^{0,2}=0$.
Consequently the local model of the extended moduli space
$\EM$ is given by $f^{-1}(0)$ where
\eqn\Kimy{
\eqalign{
f: \BH^{0,1}\oplus \BH^{3,0}&\rightarrow \widetilde{\mathbf\BH}^{2,0},\cr
(\a, \b)&\rightarrow \L(H(\a\wedge \b)).
}
}

\end{document}